Research

**Author for correspondence:**
Aditya Gulati
e-mail: aditya@ellisalicante.org

# What is beautiful is still good: the attractiveness halo effect in the era of beauty filters

Aditya Gulati[1,2], Marina Martínez-Garcia[3], Daniel Fernández[4], Miguel Angel Lozano[2], Bruno Lepri[5] and Nuria Oliver[1]

[1]ELLIS Alicante, Alicante, Spain
[2]University of Alicante, Alicante, Spain
[3]Universitat Jaume I de Castellon, Castellon, Spain
[4]Universitat Politècnica de Catalunya · BarcelonaTech, Barcelona, Spain
[5]Fondazione Bruno Kessler, Trento, Italy

AG, 0000-0002-0356-2987; MM-G, 0000-0002-2228-4396; DF, 0000-0003-0012-2094; MAL, 0000-0002-4757-5587; BL, 0000-0003-1275-2333; NO, 0000-0001-5985-691X

The impact of cognitive biases on decision-making in the digital world remains under-explored despite its well-documented effects in physical contexts. This paper addresses this gap by investigating the attractiveness halo effect using AI-based beauty filters. We conduct a large-scale online user study involving 2748 participants who rated facial images from a diverse set of 462 distinct individuals in two conditions: original and attractive after applying a beauty filter. Our study reveals that the *same* individuals receive statistically significantly higher ratings of attractiveness and other traits, such as intelligence and trustworthiness, in the attractive condition. We also study the impact of age, gender and ethnicity and identify a weakening of the halo effect in the beautified condition, resolving conflicting findings from the literature and suggesting that filters could mitigate this cognitive bias. Finally, our findings raise ethical concerns regarding the use of beauty filters.

## 1. Introduction

Beauty matters, even when we know that physical attractiveness is not correlated with other measurable traits, such as intelligence [1–3]. In fact, decades of research in several disciplines— including sociology, psychology, behavioural economics and organizational science—has found that perceptions of attractiveness profoundly impact the social judgements that we make: human beings are positively biased towards individuals who are perceived as physically attractive.

Due to this cognitive bias, known as the *attractiveness halo effect*, physically attractive people are considered to be more intelligent [4–6], happier [7,8], more trustworthy [9], more sociable and sexually warmer [10], better adjusted [11] and generally more successful in life [4], when compared with less physically attractive individuals. This halo effect has an impact on consequential aspects of our lives, as attractive individuals are thought to be better students [12] or politicians [13], more qualified for jobs [14,15], and are more likely to receive promotions, higher salaries [16,17] or more lenient judicial sentences [18,19] than less attractive people.

However, these findings have been generally obtained by means of small user studies where study participants provided judgements of a typically small sample of face images with limited diversity. Hence, questions arise regarding the generalization of the attractiveness halo effect from different perspectives.

First, concerning the ethnicity of the stimuli and the human evaluators, Albright *et al*. [20] found cross-cultural agreement in the judgements provided to western and non-western faces. However, more recent research reported a cross-cultural variation [21] and hence did not corroborate previous results. To shed light on this issue, Batres and Shiramizu [22] carried out a large-scale study that examined the attractiveness halo effect across 45 countries in 11 world regions and on a diverse set of faces from four ethnicities. Their results showed that attractiveness correlated positively with most of the socially desirable personality traits—such as being more confident, emotionally stable, intelligent, responsible, sociable and trustworthy. Hence, according to this study, the attractiveness halo effect would generalize to diverse stimuli and human evaluators. Related work by Gabrieli *et al*. [23] found that the attractiveness halo effect regarding trustworthiness is only influenced by the age of the presented faces, but not by their gender or ethnicity. Similarly, Kunst *et al.* [24] reported mixed results regarding the impact of ethnicity on the attractiveness halo effect in the context of hireability. Therefore, the evidence in this regard is inconsistent and additional research would be needed to shed light on this matter.

The second perspective relates to the interaction between the gender of the stimuli and the gender of the human evaluators. Early work by Dion *et al*. [4] did not report any significant interactions between the gender of the human evaluators and the gender of the stimulus regarding the existence of the attractiveness halo effect. However, later research reported a stronger attractiveness halo effect towards opposite-gender individuals [25]. In fact, several studies only included male raters of female faces (e.g. [26,27]) or female raters of male faces [28]. In a study with both male and female raters and stimuli, Kunst *et al*. [24] reported a significant interaction of gender, attractiveness and competence *only* when male participants rated the competence of female applicants in a hiring scenario. Again, there is mixed evidence in this regard.

The third perspective concerns the existence of this cognitive bias on the *same individual* in two conditions: original and attractive. Would the same person be perceived as having higher levels of socially desirable attributes—such as intelligence, trustworthiness or sociability—simply by improving their physical appearance?

Cosmetics are a popular tool to alter appearance and their use has been shown to increase perceptions of attractiveness [29–36]. Make-up has been reported to increase skin evenness [32] and facial contrast, which in turn leads to a perception of increased femininity and attractiveness [29]. Further literature has studied how varying levels of make-up impact perceived attractiveness [33]. While some research found that light make-up is preferred to heavy make-up [34] and others reported the opposite effect [31], faces with make-up applied to them were consistently rated as more attractive than those without make-up. Thus, the application of make-up has been used in the literature to study the attractiveness halo effect in two conditions: original and attractive. By means of user studies with psychology students and a very small set of stimuli in two conditions (original and attractive), several authors reported that the attractive condition evoked more social reinforcement and enhanced popularity ratings [26,27], and higher levels of competence, professionalism, assertiveness and ability to provide support [37]. However, others reported no statistically significant differences in the attribution of socially desirable characteristics among subjects in the original and attractive conditions [38,39]. Furthermore, these studies involved opposite-gender pairs where male participants—mainly recruited from universities—evaluated images or videos of female confederates without (original) or with (attractive) make-up applied. While insightful, these studies are difficult to scale up since it is costly to physically apply make-up to a large number of stimuli. Despite recent work showing that make-up increases perceived attractiveness in male faces as well [30], the application of make-up could create a gender asymmetry as make-up is socially more acceptable when applied to female than to male faces in many cultures [40] and the improvements in attractiveness that can be achieved as a result of applying make-up are limited.

In sum, the literature suggests that the *what is beautiful is good* notion [4] may be oversimplified, supporting the need for further research to better understand this phenomenon. Moreover, there is evidence that increased perceptions of physical attractiveness also lead to increased perceptions of socially undesirable traits, such as vanity [41–43], materialism and sexual permissiveness [44]. Nonetheless, the research presented in this paper focuses on the attractiveness halo effect related to perceptions of socially desirable attributes, namely intelligence, trustworthiness, sociability and happiness. A detailed discussion for this choice and the associated limitations can be found in §3.

In addition to shedding light on these open questions, we expand the scope of the study of this cognitive bias from the physical to the digital world. The attractiveness halo effect acquires a new relevance in the digital space, particularly as human-to-human communication is frequently mediated by technology and artificial intelligence (AI) tools are increasingly used to make assessments about humans, to interact with us via e.g. chatbots and to create enhanced digital versions of ourselves. Beauty filters are an example of such a tool, which aim to *beautify* the face of the person by applying complex transformations to the face that go beyond what make-up can achieve, including morphological changes to the eyes and eye lashes, the nose, the chin, the cheekbones and the lips, in addition to smoothing the skin, removing wrinkles and imperfections [45,46]. These filters offer a unique opportunity to study the attractiveness halo effect at scale, with diversity in the age, gender and ethnicity of the stimuli, and in a controlled scenario, because they allow the creation of *beautified* versions of the *same* individuals. While make-up has been shown to reliably increase perceptions of attractiveness [32–34], applying make-up requires technical skill, and perceptions of attractiveness differ depending upon the skill of the person applying the make-up [31]. The manipulation of beauty by means of beauty filters enables a controlled and consistent adjustment of attractiveness at a large scale while keeping the identity of the face constant [45], which is crucial for isolating the effects of perceived attractiveness from other confounding variables, such as facial identity or expression.

Furthermore, beauty filters are widely used in the digital world, and they play a significant role in shaping contemporary beauty standards and perceptions [47]. While they have been shown to profoundly impact user self-presentation—raising questions about authenticity [47], self-esteem [48–50], mental health [51,52], diversity [45] and racism [46]—there is a lack of research on how they impact perceptions of attractiveness and associated cognitive biases on the same individuals. There is also a need to study the effect of these augmented appearances on how users are perceived and judged within digital environments, both by humans and by AI algorithms. In fact, recent studies have investigated the role that beauty filters play in perceptions of trustworthiness of male stimuli [28] and of male and female stimuli in a hiring scenario [24]. Our research complements these studies by means of a large-scale user study of the attractiveness halo effect regarding four socially desirable attributes, namely intelligence, trustworthiness, sociability and happiness, with a diverse set of stimuli.

We leverage a state-of-the-art popular beauty filter applied to a diverse set of face images ($n = 462$) to create an *attractive* condition for the same individuals. Using this dataset, we perform a large-scale user study ($n = 2748$) to shed light on the conflicting findings reported in the literature regarding the attractiveness halo effect in the context of socially desirable attributes, and how different rater and stimuli characteristics, such as gender or age, impact the perception of these attributes.

Our research thus contributes to the understanding of this cognitive bias from four different perspectives: first, we study the impact of the beauty filters on the attractiveness halo effect for the *same individuals*; second, we investigate the existence of this cognitive bias on a *diverse* set of stimuli (faces); third, we analyse the role that the gender, age and ethnicity of the stimuli and the raters play regarding the attractiveness halo effect; fourth, we explore the potential of beauty filters to mitigate the existence of the attractiveness halo effect in the digital world.

## 2. Results

We report the results of analysing the responses of 2748 study participants (raters) who provided ratings on a 7-point Likert scale for seven different attributes—namely, attractiveness, intelligence, trustworthiness, sociability, happiness, femininity and unusualness—in addition to their estimation of the gender, age and ethnicity of 10 different face images (stimuli) from a pool of 924 images. A detailed description of the study procedure and design can be found in §4.3, while a summary can be seen in figure 1.

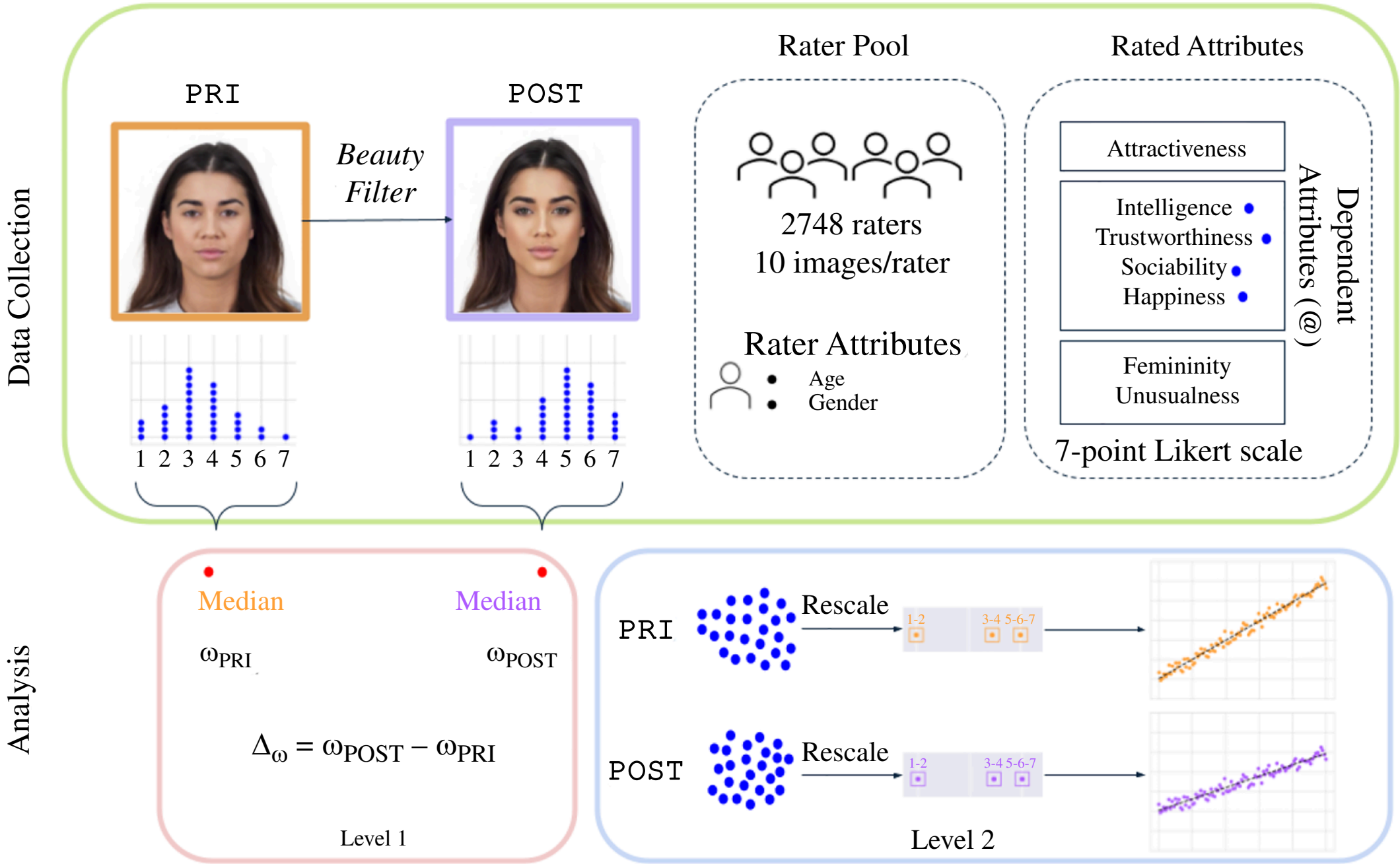

**Figure 1.** Overview of the study and the analysis of the collected data. The stimuli consist of two sets of facial images: the PRI dataset, extracted from existing datasets for research on faces [53,54] and the POST set, created by applying a state-of-the-art beauty filter to each image in the PRI dataset. Each participant ($N$ = 2748) rated 10 different images on seven attributes indicated on the top right part of the figure. Each image received ratings from at least 25 different participants. To shed light on the attractiveness halo effect, two levels of analysis were performed: (i) an *aggregate* level—depicted inside the pink box in the figure—using the medians of all the ratings received by each image, which are referred to as *centralized* ratings (.); and (ii) an *individual* level (.)—depicted inside the blue box in the figure—consisting of each rating and considering the participants' characteristics.

The images consisted of the original faces ($n$ = 462, labelled as PRI for **P**icked **R**epresentative **I**mages) and their corresponding beautified versions ($n$ = 462, labelled as POST for **PO**st **S**ocial media **T**ransform) by means of applying a state-of-the-art, popular beauty filter. No participant provided ratings on the same set of images to ensure that each participant was exposed to a diverse set of stimuli while maximizing the number of ratings provided for each face image. Furthermore, no participant rated an image corresponding to the *same* individual in both conditions (with and without the filter applied) and participants were not told that half of the images that they evaluated corresponded to the *beautified* versions of the original face images.

The reported results are structured according to two levels of analysis. Following past studies [55,56], we first compute the median value—due to the non-normality in the distribution of the values ($D$ = 0.93, $p < 0.001$, Kolmogorov–Smirnov)—of the ratings provided by the participants for each image and each attribute, which is henceforth referred to as the *centralized* score. While this level of analysis enables making pairwise comparisons between the ratings provided to the same individuals in the PRI and POST sets, it does not allow to study the variance in the ratings due to the participants. Thus, we also analyse each rating individually to include the effects of the participants' gender and age. To perform such an analysis, ordered stereotype models (OSMs) [57–59] are first applied to the ordinal responses on the 7-point Likert scales to estimate 'a new spacing among the ordinal categories dictated by the data' [59]. The raw data is then transformed according to the new scales obtained with the OSMs and we build linear mixed models to study the impact of the raters' gender and age on their responses, considering the raters as random effects. A detailed discussion of the methodology used to analyse the ratings can be found in §4.

## 2.1. Beauty filters and attractiveness

### 2.1.1. Manipulation test: do beauty filters increase attractiveness?

The same individuals were rated as significantly more attractive after applying the beauty filter than before its application ($p < 0.001$, one-sided Wilcoxon paired-rank), as reflected in figure 2*a*, which depicts the distribution of centralized attractiveness ratings for each image before and after the filter was applied.

The median increase in perceived attractiveness after beautification was 1 point on the 7-point Likert scale. There were no images where the centralized perceived attractiveness score decreased after beautification and it remained the same before/after beautification only in 3.9% (18 out of 462 images) of the cases. We conclude, thus, that the manipulation was successful as the beauty filters significantly increased the perceived attractiveness of the same individuals after beautification.

The increase in perceived attractiveness ($\Delta_{attrac} = \text{Attrac}_{\texttt{POST}} - \text{Attrac}_{\texttt{PRI}}$) due to the application of the beauty filter is negatively correlated with the initial attractiveness score of the face images (Kendall's $\tau = -0.49$, $z = -12.395$, $p < 0.001$), as reflected in figure 2*b*: the lower the initial attractiveness, the larger the benefit of applying the beauty filter.

### 2.1.2. Impact of the filters regarding the age, gender and ethnicity of the stimuli

Figure 3 depicts the centralized attractiveness scores in the original (`PRI`) and beautified (`POST`) datasets according to the age, gender and ethnicity of the stimuli. Note that we adopt the same nomenclature as the labels provided in the face datasets analysed in our study: gender is a binary variable with two values (male/female) and ethnicity can have six values (Asian/Black/Latino/White/Indian/Mixed). As explained in §4, the analyses of age and ethnicity are carried out on the images from the FACES dataset and the Chicago Faces Database (CFD) respectively, whereas the analysis of gender is performed on all the images from both datasets. The age groups are given by the FACES dataset and correspond to: Young [$19 \le age \le 31$]; Middle [$39 \le age \le 55$]; and Old [$age > 69$]. As seen in figure 3, while the age and gender of the stimuli have a clear impact on their perceived attractiveness levels, ethnicity does not seem to play a role.

More precisely, a statistically significant difference in the centralized perceived attractiveness scores depending on the age and gender of the individual was found, both in the original (`PRI`) and beautified (`POST`) versions ($p < 0.001$, Kruskal–Wallis). No statistically significant effect of ethnicity was found in either of the conditions. Images corresponding to young individuals received significantly higher ($p < 0.001$, pairwise Wilcoxon) centralized perceived attractiveness scores than those depicting middle-aged or older individuals in both the `PRI` and `POST` sets. Images depicting middle-aged individuals were considered significantly ($p < 0.001$, pairwise Wilcoxon) more attractive than those depicting older individuals only after applying the beauty filter.

The increase in perceived attractiveness ($\Delta_{attrac}$) due to the filters was also significantly different across age groups. Images of middle-aged individuals had a mean $\Delta_{attrac}$ of 1.57 points, which was significantly ($p < 0.001$, pairwise Wilcoxon) higher than the $\Delta_{attrac}$ of images corresponding to younger individuals, who had a mean increase of 1.18 points in their centralized attractiveness scores due to the application of the filters. Images depicting older individuals had a mean $\Delta_{attrac}$ of 1.38 points, which did not differ significantly from images of either younger or middle-aged individuals.

Images of females received significantly higher ($p < 0.001$, Kruskal–Wallis) perceived attractiveness ratings than images of males both before and after beautification. The mean increase in centralized attractiveness for female images ($\Delta_{attrac} = 1.53$) was higher ($p < 0.01$, Kruskal–Wallis) than that for male images ($\Delta_{attrac} = 1.34$). A similar analysis on the impact of the filters on the dependent attributes can be found in appendix G.

In addition, we study how the filters impact the perception of physical characteristics such as age, gender and ethnicity along with attributes related to physical appearance, such as perceived femininity and unusualness. These findings are reported in appendix C.

In the following sections, we focus on the attractiveness halo effect regarding four attributes that have been extensively studied in the literature: intelligence [6,22,60,61], trustworthiness [4,9,22,53,60–63], sociability [4,22,60] and happiness [8,22,53,60,62].

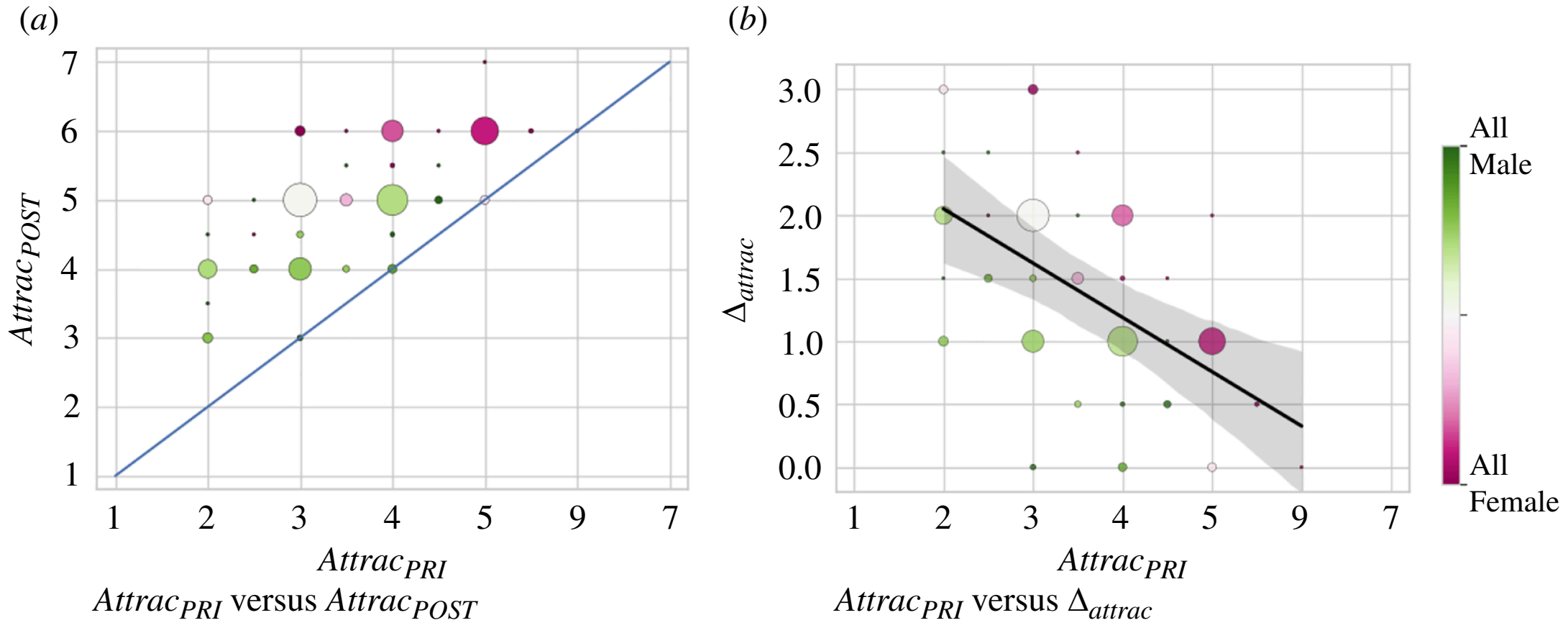

**Figure 2.** Impact of the beauty filters on perceived attractiveness. The size of the circles is proportional to the number of ratings provided for each value on the 7-point Likert scale and the colour indicates the proportion of males and females for each rating. (*a*) Pairwise comparison of perceived attractiveness before and after beautification. Observe how no image decreased its perceived attractiveness ratings after beautification and how the highest perceived attractiveness ratings tend to correspond to females. (*b*) Increase in perceived attractiveness ($\Delta_{attrac}$) after the application of the beauty filter versus the initial levels of attractiveness. Shading corresponds to the 95% confidence interval. The higher the original perceived attractiveness, the lower the increase in attractiveness after applying the filter.

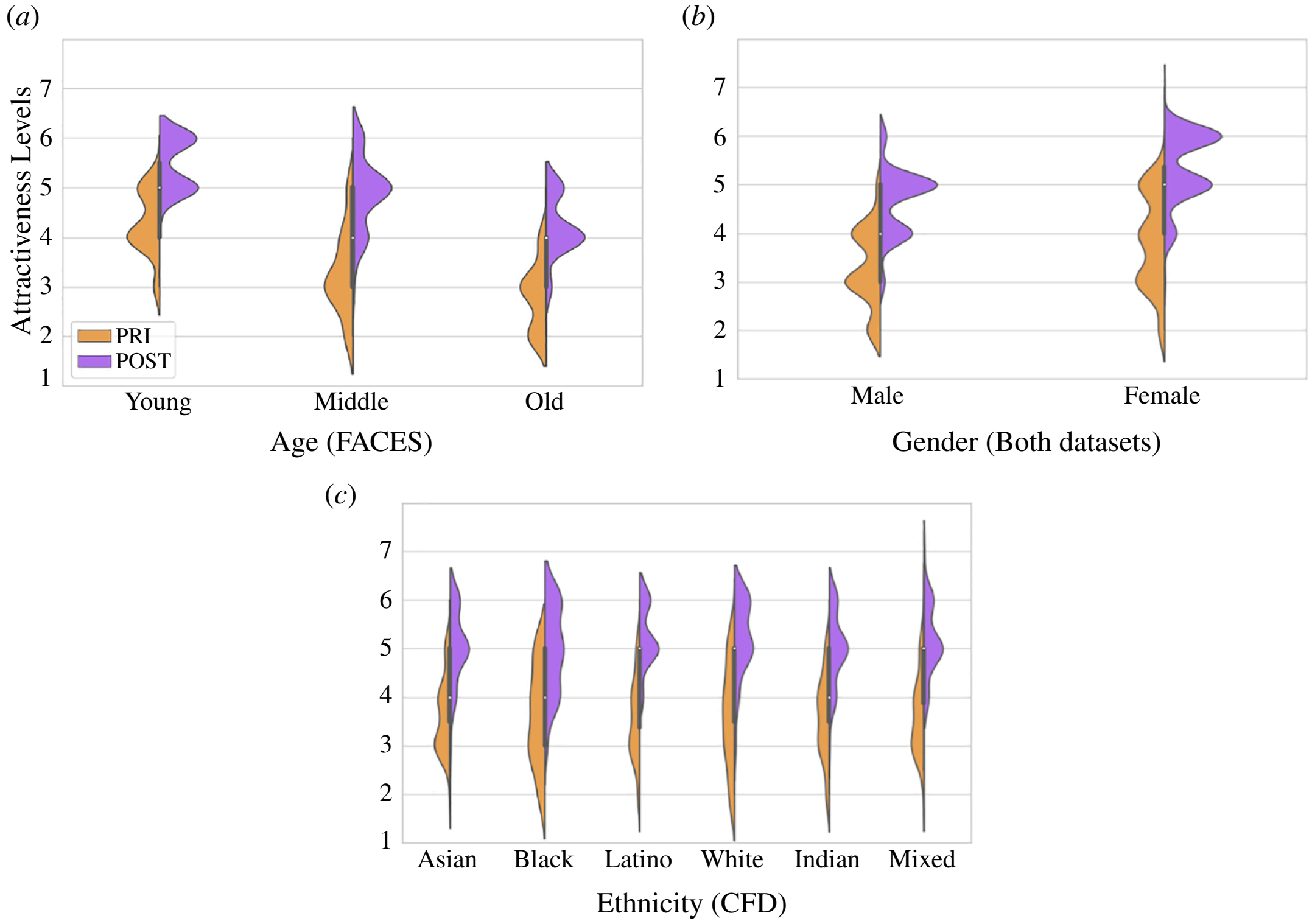

**Figure 3.** Distribution of the median ratings of perceived attractiveness of the original (`PRI`, in orange) and beautified (`POST`, in purple) face images when varying the age (*a*), gender (*b*) and ethnicity (*c*) of the stimuli. Note that the age and ethnicity results are computed on the FACES and CFD datasets, respectively, whereas the gender results are based on the analysis of both datasets. Regarding age, the younger the individual, the higher their perceived attractiveness ratings ($p < 0.001$, pairwise Wilcoxon). With respect to gender, female faces receive higher attractiveness ratings than male faces ($p < 0.001$, Kruskal–Wallis). No statistically significant difference was found in the attractiveness levels depending on the ethnicity of the stimuli both before and after beautification.

**Table 1.** Parameters of the linear model $\omega = \beta_0 + \beta_1 Attrac + \epsilon$ for each dependent variable $\omega$ on the PRI and POST sets independently. A larger absolute value of the intercept $\beta_0$ in the POST set indicates that the value of the perceived attribute increases after applying a beauty filter. A smaller absolute value of $\beta_1$ in the POST set reflects a weaker halo effect after beautification.

| dependent attribute ($\omega$) | PRI | | | POST | | |
|---|---|---|---|---|---|---|
| | $\beta_0$ | $\beta_1$ | $R^2$ | $\beta_0$ | $\beta_1$ | $R^2$ |
| intelligence | 3.18*** | 0.30*** | 0.327 | 4.11*** | 0.12*** | 0.036 |
| trustworthiness | 3.34*** | 0.20*** | 0.181 | 3.50*** | 0.17*** | 0.069 |
| sociability | 2.56*** | 0.39*** | 0.363 | 2.78*** | 0.38*** | 0.321 |
| happiness | 2.08*** | 0.39*** | 0.261 | 2.47*** | 0.35*** | 0.186 |

## 2.2. Beauty filters and the attractiveness halo effect

Statistically significant differences were found in the centralized scores of the four dependent variables of interest (intelligence, trustworthiness, sociability and happiness) between the original (PRI) images and their beautified (POST) versions ($p < 0.001$, one-sided Wilcoxon paired-rank). Images of the *same individuals* received higher scores on all attributes after beautification, as depicted in table 6 (appendix J). Thus, the *same individuals* were perceived not only as more attractive, but also as more intelligent, trustworthy, sociable and happy after applying a beauty filter, providing evidence that supports the existence of the attractiveness halo effect.

Linear models, depicted in table 1, of the centralized score for each dependent variable ($\omega$) as a function of the centralized score of perceived attractiveness for each image ($\omega = \beta_0 + \beta_1 Attrac + \epsilon$) reveal a significant effect ($p < 0.001$)[1] of perceived attractiveness on all dependent variables both before (PRI) and after (POST) beautification. The positive and significant $\beta_1$ for all attributes on the PRI and POST sets supports the existence of the halo effect and is in line with past work that studied this effect using different subjects in two conditions: original and attractive [26,27,37]. *Intelligence* exhibits the largest decrease in $\beta_1$ after beautification, reflecting a weaker halo effect. There is a significant decrease in the goodness-of-fit of the model ($R^2$) for intelligence (approx. 90%) and trustworthiness (approx. 60%). We discuss the implications of these findings in §3.

## 2.3. Impact of the raters on the attractiveness halo effect

The centralized ratings allowed performing pairwise comparative analyses between the images in the PRI and POST datasets. However, aggregating the scores by their medians masks the impact of the raters' attributes, such as their age and gender, on the perceptions of attractiveness and the attractiveness halo effect. In this section, we report the results when analysing each rating individually to consider the role of different rater characteristics in the perception of the dependent attributes, and the halo effect.

To leverage the individual ratings, the collected ordinal ratings were first transformed into a continuous variable using the OSM [57–59]. For the data in the PRI and POST datasets independently, we then built linear mixed models of perceived attractiveness (equation (4.5)) and of each of the dependent variables using attractiveness and the rater's characteristics (age and gender) and their interactions as independent variables (equation (4.6)). A detailed discussion motivating this modelling choice can be found in §4.5. The new scales for attractiveness and the dependent attributes ($\omega$) computed by the OSM can be found in appendix B.

Note that pairwise comparisons between images in the PRI and the POST datasets—as was done with the centralized scores—are not appropriate for two reasons. First, since no participant rated the same image both before and after beautification, it is not possible to generate any logical pairs. Second, the OSM is computed independently on the PRI and POST sets as the goal of this part of the analysis is understanding the impact of different rater attributes on perceptions with and without the filters.

[1]In the figures and tables, we use the standard star notation to represent the $p$-values, i.e. *** : < 0.001, ** : < 0.01 and * : < 0.05

**Table 2.** Significance levels (*** $p < 0.001$; ** $p < 0.01$) and magnitudes of the $\beta$'s in the linear mixed models built to measure the impact of the rater's and stimulus's age and gender on the attractiveness halo effect. The shading in each cell corresponds to the absolute value and sign of the corresponding $\beta$ on normalized data in order to compare their effect across different variables. $\beta_1$:$Attrac_I$, $\beta_2$:$Gender_I$, $\beta_3$:$Age_I$, $\beta_4$:$Gender_R$, $\beta_5$:$Age_R$, $\beta_6$:$Gen_I \cdot Gen_R$, $\beta_7$:$Age_I \cdot Age_R$. Note how perceived attractiveness is the strongest predictor both before and after beautification. After beautification, other variables play a role given the decreased predictive power of attractiveness, details of which can be found in appendix M.

| $\omega$ | $\beta_0$ | | $\beta_1$ | | $\beta_2$ | | $\beta_3$ | | $\beta_4$ | | $\beta_5$ | | $\beta_6$ | | $\beta_7$ | |
|---|---|---|---|---|---|---|---|---|---|---|---|---|---|---|---|---|
| | PRI | POST | PRI | POST | PRI | POST | PRI | POST | PRI | POST | PRI | POST | PRI | POST | PRI | POST |
| Attractiveness | *** | *** | x | x | *** | *** | *** | *** | *** | *** | | | *** | | *** | *** |
| Intelligence | *** | *** | *** | *** | *** | *** | ** | *** | | ** | | *** | | | | *** |
| Trustworthiness | *** | *** | *** | *** | *** | *** | *** | *** | | *** | | *** | | ** | | |
| Sociability | *** | *** | *** | *** | | *** | | *** | *** | | | | | | | |
| Happiness | *** | *** | *** | *** | ** | *** | | | *** | | | *** | | ** | | |

This leads to different scales for the attributes between the PRI and POST sets due to which pairwise comparisons are not appropriate.

Table 2 summarizes the $\beta_i$s and associated $p$-values for each of the linear mixed models. Note how all $\beta_0$ and $\beta_1$ are significant ($p < 0.001$) for perceived attractiveness and the dependent variables both before and after beautification. Perceived attractiveness ($\beta_1$) is the strongest predictor of the dependent variables, yet its predictive power decreases in the models built with data after beautification (see appendix M for a detailed analysis). As a consequence, there are other factors that play a more significant role after beautification. The colours in table 2 represent the values of the $\beta_i$s on a normalized scale to allow for an easier comparison across models. Using these models, we analyse next the impact on the attractiveness halo effect of the rater's age and gender, and their interactions with the age and gender of the stimuli.

### 2.3.1. Impact of the rater's age

Regarding attractiveness, the perceived age of the stimulus ($\beta_3$) is negatively correlated ($p < 0.001$) both before and after beautification, as has been extensively reported in the literature [64–66]. Conversely, the rater's age ($\beta_5$) does not exhibit a significant correlation with perceived attractiveness, which is aligned with previous research [67,68] and in contradiction with what other authors have reported [69]. Furthermore, we observe a significant ($p < 0.001$) positive correlation in the interaction between the perceived age of the stimulus and the rater ($\beta_7$), in concordance with the literature [66].

With respect to the dependent variables, the rater's age has a statistically significant positive correlation ($p < 0.001$) with perceived intelligence, trustworthiness and happiness only after beautification. There is no statistically significant impact of the rater's age on any of the dependent variables before beautification. Interestingly, the interaction between the rater's and stimulus's age is only significant for perceptions of intelligence after beautification.

### 2.3.2. Impact of rater's gender

As with age, the models represented by equations (4.5) and (4.6) consider the impact of the rater's gender ($\beta_4$), the stimulus's gender ($\beta_2$) and their interaction ($\beta_6$). Note that the significance levels reported for $\beta_6$ in table 2 correspond only to the interaction term of male raters rating male images since females were encoded as 0. Thus, we report the estimated marginal means [70,71] for each (image gender, rater gender) pair. Figure 4 depicts the estimated marginal means for attractiveness and the four dependent attributes for all (image gender, rater gender) pairs in the PRI and POST datasets.

Before beautification, both male and female raters provide significantly different scores of attractiveness ($p < 0.001$), trustworthiness ($p < 0.001$) and intelligence ($p < 0.01$ for female raters, $p < 0.05$ for male raters) to images of males and females. Images of females receive significantly higher: (i) sociability ($p < 0.001$) scores from male raters than from female raters; and (ii) attractiveness ($p < 0.001$) scores from female raters than from male raters before beautification. There are no statistically significant differences ($p < 0.001$) in the scores assigned to images of males regarding all attributes when rated by both male and female raters.

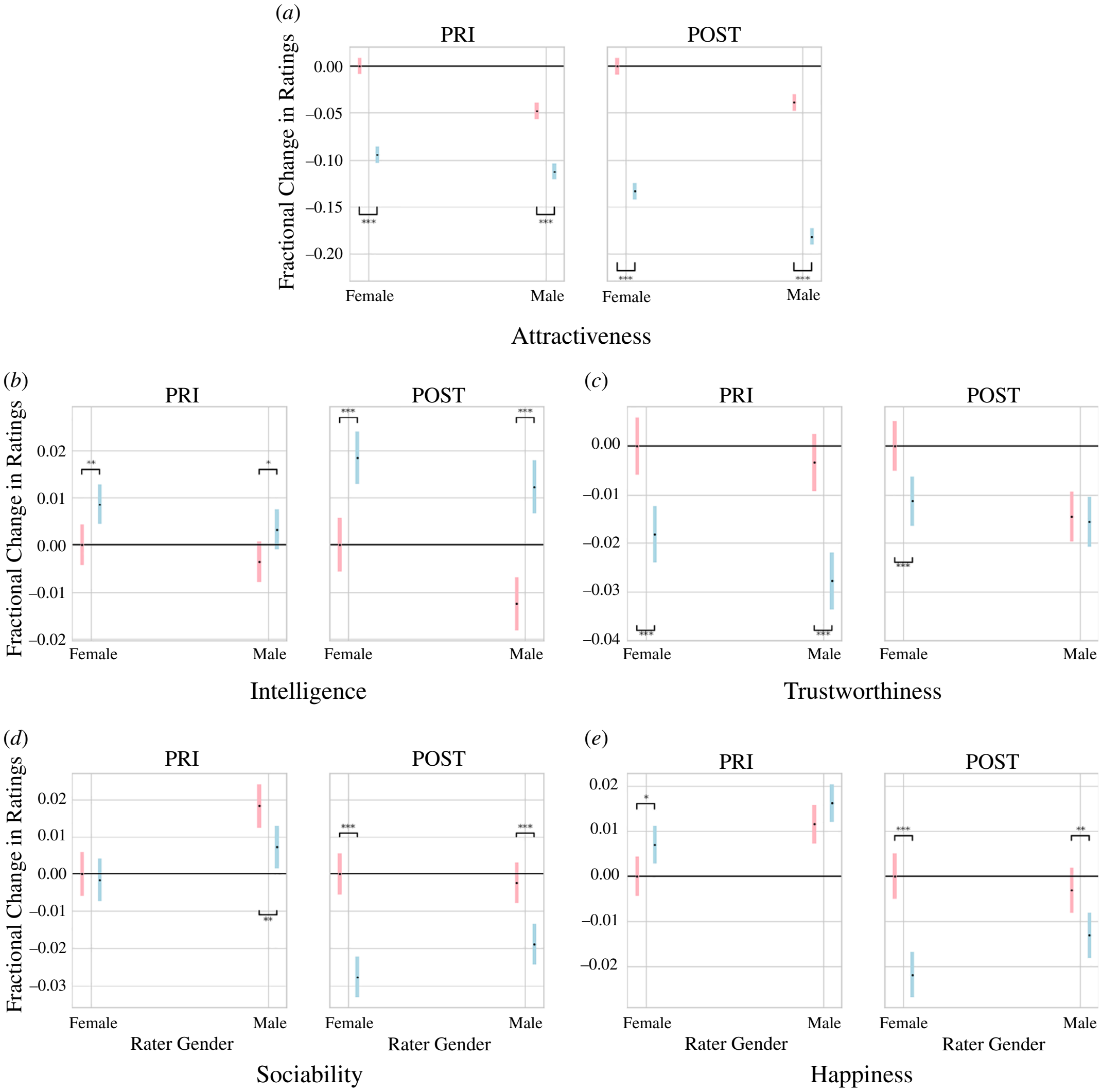

**Figure 4.** Impact of rater's and stimulus's gender on attractiveness and the dependent variables in the PRI and POST datasets. The *x*-axis represents the gender of the rater and the colours represent the gender of the stimulus (pink [.] for images of females and blue [.] for images of males). The length of the bars corresponds to the 95% confidence interval of the estimated marginal mean (EMM) [70,71]. The *y*-axis depicts the relative change in the EMM from the EMM of female stimuli rated by female participants. Details on how these values were computed can be found in appendix N.

After beautification, female raters provide significantly different ($p < 0.001$) ratings to images of males and females on all attributes, whereas male raters provide significantly different ($p < 0.001$) scores to images of males and females only on perceived attractiveness, intelligence and sociability but not on trustworthiness and happiness. While images of males received comparable scores on all attributes in the PRI dataset, there are statistically significant differences ($p < 0.001$) in the perceived attractiveness of images of males by male and female raters after beautification, with male raters providing lower scores to images of males than female raters. Images of females are also given significantly lower attractiveness ($p < 0.001$) scores by male raters than by female raters, as observed in the PRI dataset. Additionally, images of females are given significantly lower trustworthiness scores by male raters, even though male and female raters provided similar trustworthiness scores to females in the PRI dataset. The opposite impact of the filters is seen regarding sociability, with no significant difference observed in the ratings received by images of females despite there being a significant difference before beautification.

Even though images of females were given higher scores of perceived attractiveness ($p < 0.001$) than images of males by both male and female raters, they were given lower scores of intelligence than images of males, particularly after beautification ($p < 0.001$). This finding suggests the existence of a gender bias in perceptions of intelligence [3,6]. Gender has also been found to play a significant role in the perception of related attributes such as competence and hireability [24,72–75]. The implications of this finding are discussed in §3.

Figure 4 also provides insights into the impact of the filters on male and female raters. The gap between the ratings given to images depicting males versus females by male and female raters before and after beautification notably increases when judging attractiveness, intelligence, sociability and happiness and decreases when judging trustworthiness. Moreover, the gender differences in attractiveness, intelligence and trustworthiness ratings change significantly more after beautification for male raters, whereas a similar effect is observed for sociability and happiness for female raters. Finally, trustworthiness is the only dependent variable where the gender differences in the scores provided to images of males and females by male and female raters decrease after beautification. Table 5 quantifies the percentage change in ratings for different dependent attributes depending on the gender of the rater.

These findings suggest that judgements made by male raters on attractiveness, intelligence and trustworthiness are more sensitive to the filters when compared with the judgements by female raters. Conversely, female raters tend to be more sensitive to the beauty filters than male raters when providing judgements of sociability and happiness. Implications of these findings are discussed in §3.

### 2.4. Do beauty filters mitigate the attractiveness halo effect?

Beauty filters increase the perceived attractiveness scores for almost all individuals indicating that they shift the distribution of perceived attractiveness to the right on the 7-point Likert scale. Additionally, they have a greater impact on individuals who received low scores of perceived attractiveness before beautification (figure 2*b*). This leads to beauty filters narrowing the spread of perceived attractiveness ratings ($p < 0.001$, Levene's [76]), thereby reducing their influence as a factor to impact the perception of other attributes, such as intelligence. Thus, beauty filters could potentially mitigate the halo effect.

The linear models in table 1 reflect a decrease in the value of $\beta_1$ and $R^2$ of the linear models after beautification, particularly for intelligence and trustworthiness, supporting the hypothesis of a mitigation of the halo effect for these attributes. We postulate the existence of a saturation effect, i.e. beyond a certain level of perceived attractiveness, there is a significant reduction in the impact that attractiveness has on the dependent variables.

Figure 5 depicts the relationship between perceived attractiveness and the dependent variables after rescaling the data according to the OSMs both before (*a*) and after (*b*) beautification. In the case of intelligence, we observe a clear saturation effect in the `PRI` dataset, and a similar effect is observed for trustworthiness in the `POST` dataset, where the slope of the linear mixed model of trustworthiness as a function of perceived attractiveness decreases as attractiveness increases, especially when compared with sociability and happiness. Detailed statistical analyses supporting this saturation effect can be found in appendix K.

These findings suggest that the filters' capacity to enhance attractiveness, rather than their ability to reduce attractiveness variation, is the main factor in reducing the halo effect observed in certain attributes. Implications of this saturation effect are discussed next.

## 3. Discussion

In this study we have collected human feedback of a large-scale, diverse dataset of face images of the same individuals in unattractive (original) and attractive (beautified) conditions by means of applying a digital beauty filter. While personal preferences arguably play a role in perceptions of attractiveness [77–79], we obtain irrefutable evidence that AI-based beauty filters increase the perceptions of attractiveness for almost all individuals, regardless of their gender, age and race. The centralized perceived attractiveness score increased for 96.1% of the individuals after beautification and remained unchanged for the rest.

Few studies [24,26,27,37,39] have investigated the presence of the attractiveness halo effect on the *same individual* by creating two conditions: an attractive and unattractive setting for the same person.

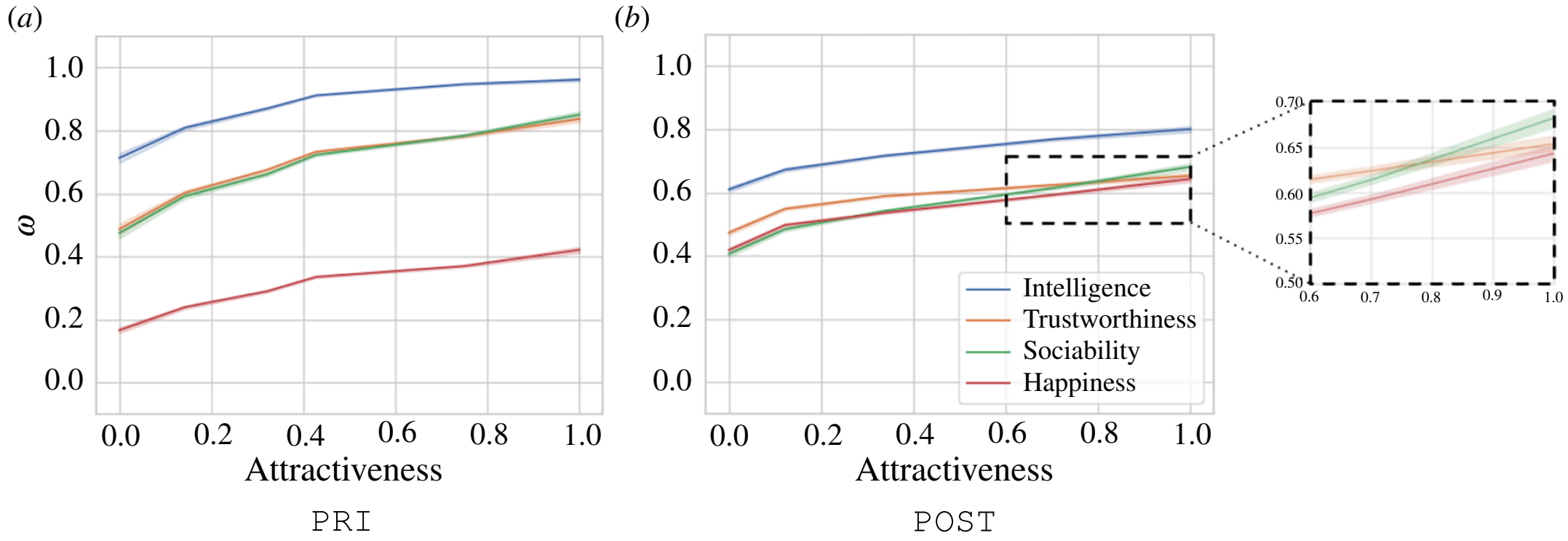

**Figure 5.** A visual representation of the relationship between perceived attractiveness and the dependent attributes after rescaling with the ordered stereotype model. The scales here have been normalized for ease of representation. Note how intelligence shows a much stronger saturation effect than the other dependent attributes in the `PRI` dataset. In the `POST` dataset, both intelligence and trustworthiness exhibit a saturation effect.

The attractive condition was typically achieved by enhancing or beautifying the appearance of the individual to be rated by means of professional lighting, fashionable clothing and hair style, the application of make-up and/or, more recently, digital beauty filters. The results of previous studies have been mixed. The diversity of our stimuli, our large sample and the ability to apply a consistent transformation to increase attractiveness by means of beauty filters provide robust data on this matter. Contrary to previous works [38,39] and supportive of others [26,27,37], we find strong evidence of the existence of the halo effect both before and after beautification for the four dependent variables of interest (see table 1). Furthermore, beauty filters impact the attractiveness halo effect differently, depending on the attribute: while still significant for all dependent variables, the effect weakens after beautification for intelligence and trustworthiness (table 1), suggesting that beauty filters could be used to mitigate the attractiveness halo effect regarding these two attributes due to the increase in the attractiveness levels after applying the filter. In fact, the mean value for perceived attractiveness increased from 3.57 in the `PRI` dataset to 5.01 in the `POST` dataset. As a result, while only approximately 17% of the faces in the `PRI` dataset were rated as having an attractiveness level greater than or equal to 5 (with 4 being the neutral point on the scale), this percentage increased to approximately 75% after beautification. Furthermore, the distribution of attractiveness values decreased its variance after beautification, dropping from 0.83 in the `PRI` dataset to 0.60 in the `POST` dataset. Additionally, we identify a negative correlation between the original levels of attractiveness and the increase in attractiveness, such that the larger the perceived attractiveness of the original image, the smaller its increase in attractiveness due to the application of the filter (figure 2*b*).

Additional analyses revealed that the relationship between attractiveness and the dependent variables is nonlinear such that it saturates after a certain level of perceived attractiveness is surpassed (§2.4). The strength of the saturation is different for each dependent variable, being the strongest for intelligence and trustworthiness. The difference in strength of the saturation effect is consistent with previous work that has shown that the strength of the attractiveness halo effect is trait selective [11,44]. The halo effect in fact is not only trait selective in strength but also in direction. While the traditional 'what is beautiful is good' notion [4] would suggest that increased attractiveness leads to increased positive impressions, studies have shown that an increase in attractiveness is also correlated with an increase in the perception of certain negative traits, such as vanity [42,43], egotism [41], materialism and sexual permissiveness [44]. Scholars have tried to identify a functional basis of attributes that are used to evaluate faces [60], yet generalizing the findings about the halo effect to any trait is non-trivial. Since our study did not include negative attributes, it is yet unclear to which degree a potential saturation effect would be present in these situations.

However, the identified saturation effect provides a unifying explanation for several inconsistent findings reported in the literature regarding the existence [38,39] and strength [11,41,80] of the attractiveness halo effect. For example, Timmerman and Hewitt [39] did not find evidence of the attractiveness halo effect based on photographs of two female models from the Cosmopolitan magazine before and after professional make-up was applied. A manipulation test concluded there was a significant change in perceived attractiveness, yet no significant changes in the perceptions of their dependent attributes (including intelligence) were found. Based on our research, their findings could be an instance of the saturation effect, especially if the stimuli were highly attractive women as could be the case given that they were selected from the Cosmopolitan fashion magazine.

Note that previous work has suggested that the halo effect and trait sensitivity could be interpreted as a stereotype effect [43]. In this regard, the saturation effect could be explained by the application of different stereotypes depending on the attractiveness levels of the stimuli. As discussed below, we find evidence of the existence of a gender bias when judging the intelligence of female stimuli, which could correspond to the application of a different stereotype for highly attractive females. However, our study design does not enable the establishment of a causal link between stereotype formation and the observed saturation effect. We leave to future work the study of such a link.

Concerning the existence of the attractiveness halo effect with a diverse set of stimuli according to ethnicity, age and gender, there is mixed evidence in the literature, which our study contributes to disambiguating [20,22–24,81].

In terms of ethnicity, our findings contradict previous work that reports that the attractiveness halo effect does not generalize when evaluating members of an ethnicity other than their own [25]. Conversely, we find strong evidence of the existence of the attractiveness halo effect for all stimuli across ethnicities, even when evaluated by participants of a different ethnicity. Therefore, we conclude that the attractiveness halo effect does generalize when evaluating members of an ethnicity other than their own, in alignment with the findings reported in [22].

The age of the rater did not have a statistically significant effect on perceptions of attractiveness but had a statistically significant positive effect on perceived intelligence, trustworthiness and happiness after beautification. This finding complements previous work that studied the existence of the attractiveness halo effect and the babyface stereotype in young and older adult raters [61]. The authors reported that older adults are as vulnerable as young adults to the attractiveness halo effect: they judged more attractive people as more competent and healthy, and less hostile and untrustworthy, corroborating previous research on young adults [11,82]. In our work, we also find that the age of the stimulus matters. In terms of perceived attractiveness, both before and after beautification young individuals were rated as significantly more attractive than middle-aged and older individuals, in accordance with prior work [64–66]. The negative and significant correlation between perceived intelligence, trustworthiness and age (particularly after beautification) suggests that the older the stimulus, the more intelligent and trustworthy it is perceived. This finding is aligned with previous literature that has reported on the *wisdom bias* [83] but contradicts recent work on trustworthiness and age [84]. Conversely, youth is positively correlated with sociability, especially after beautification, which is supportive of previous research [85].

Regarding gender, our results unveil novel interactions between the gender of the stimulus, the gender of the rater and the attractiveness halo effect, both when rating same- and opposite-gender stimuli. Images of females were rated as significantly more attractive than males, in alignment with previous research [65,67,86] and in contradiction to others [66]. Both female and male raters provided higher ratings of attractiveness to images of females before ($p < 0.001$) and after ($p < 0.001$) beautification, with a widening gap between genders after beautification, especially for male raters (figure 4*a*). Conversely, participants considered males to be more intelligent than females, particularly after beautification ($p < 0.001$), also with a widening gap between genders (figure 4*b*). Therefore, we conclude that the gender of the stimulus plays a stronger role in impacting the perceptions of intelligence than perceived attractiveness given that images of females were rated as more attractive than those of males. This finding could be explained by the application of a different stereotype to highly attractive females. We leave to future work the exploration of this potential reason for this finding.

Concerning opposite-gender effects, our findings contribute with nuanced evidence of what has been previously reported [25–28]: we observe statistically significant ($p < 0.001$) differences in the ratings provided by both female and male raters to images of opposite gender individuals for perceived attractiveness, intelligence and trustworthiness both before and after beautification, and for sociability and happiness only after beautification. As described in the previous paragraph, male stimuli are perceived as more intelligent than female stimuli both by male and female raters, with a widening gap between genders after beautification such that female stimuli are perceived as *less intelligent* on average by male raters after beautification than before applying the filter. With respect to trustworthiness, the images of females in the PRI dataset were considered to be more trustworthy by both male ($p < 0.001$) and female raters ($p < 0.001$), yet male raters considered images of males and females to have similar levels of trustworthiness after beautification. Sociability and happiness behave similarly and exhibit a widening gap between genders: men are perceived as less sociable and happy than women after beautification and especially when judged by women. In sum, we observe several and novel significant interactions between the gender of the stimulus and the gender of the human evaluators, contradicting early work that reported a lack of such an interaction [4].

The findings regarding perceived intelligence suggest that there exists a stronger gender bias than the attractiveness halo effect [87,88] and underscores deeper cultural attitudes and stereotypes surrounding gender roles and expectations [89]. Moreover, our results are supportive of previously reported examples of gender-based discrimination and the challenges faced by women in various spheres of life, including education and professional opportunities [90–93]. The perpetuation of such stereotypes can contribute to systemic inequalities and hinder the advancement of women in society [94,95]. Given the prevalence in the use of beauty filters by young females—90% of women aged between 18 and 30 report using beauty filters before posting selfies on social media [96]—our findings raise additional concerns about the potential negative impact of beauty filters on young women, a group that has been shown to be more susceptibility to body dissatisfaction [97,98]. Frequent use of beauty filters has already been found to lead to anxiety and depression, reduced self-esteem, body dysmorphia, an increase of plastic surgery, feelings of inadequacy and increased pressure to conform to unrealistic beauty standards [47,50,52,96,99–102]. Our research adds a new dimension to the harmful consequences of using beauty filters by empirically demonstrating that females are perceived by men as *less intelligent* after the application of the filters. Moreover, their use raises questions about authenticity and honesty as they alter the appearance of users, often presenting an idealized or unrealistic version of themselves. This alteration can blur the line between reality and artificiality, leading to questions about what is genuinely authentic in digital self-representation [50]. The discrepancy between real and filtered images can undermine personal authenticity and contribute to a false sense of identity [49]. There is therefore a need for transparency and ethical guidelines surrounding the use of beauty filters, especially in contexts where individuals may be influenced in their decision-making by filtered images without their knowledge.

Our study, however, is not without its limitations. First, while we included a large and diverse set of stimuli judged by over 2700 participants, the participants lacked geographic—and thereby ethnic—diversity because they consisted of predominantly white individuals from the USA and the UK. As described in §4.1, participants had to be native English speakers to qualify for the study as it was designed and deployed in English. Nonetheless, previous work has reported that the attractiveness halo effect generalizes across countries [22]. Second, we report findings at an aggregate level. A per-rater level analysis, while interesting, is not possible on our collected dataset because each participant was exposed to a different set of images to maximize diversity. Third, we do not explore the impact of different beauty filters on the halo effect, but previous work has reported that popular beauty filters perform similar transformations to the faces [45]. We also do not study the potential perceived differences between real and beautified faces for the same level of attractiveness. While certainly interesting, these questions are out of the scope of this study and hence we leave them to future work. Fourth, we do not study the relationship between attractiveness and socially undesirable attributes, such as vanity or materialism. We focus on socially desirable attributes instead because adding negative characteristics to our study would entail flipping the Likert scale for those attributes, which could have led to confusion in the participants. We leave to future work the investigation of the 'dark side' of beauty. Fifth, a related but unexplored phenomenon in our research is the halo update effect [43], according to which raters update their judgements over time, especially when presented with new information [103–105]. Since participants in our study were not presented with images of the same individual before and after beautification, we leave the investigation of the halo update effect for future research. Finally, photographs provide only a static, two-dimensional representation of individuals, lacking the multi-dimensional and dynamic nature of interactions in the physical world, where attractiveness perceptions can be influenced by factors beyond facial appearance. Hence, our findings might not generalize to real-world scenarios where attractiveness perceptions interact with other factors, such as situational dynamics, personality and social context. Nonetheless, most of the previous work that has studied this cognitive bias has adopted a similar methodology to ours [6,22,53,54,61,106–110] and faces play a significant role in our judgements of the attributes studied in this work [4,6,8,9,22,53,54,60–63].

# 4. Methods

The user study was pre-registered in the Open Science Foundation registry[2] and was approved by the Ethics Board of the University of Alicante.

## 4.1. Study participants

The study participants were recruited via the Prolific participant recruitment platform. The target sample were adults with unimpaired vision who were English native speakers. Given the purpose of the study, participants were required to be neurotypical, without any mental health condition or dyslexia and to have an approval rate of at least 85% in past studies on Prolific. The sample ($n$ = 2748) was gender balanced: 1375 men and 1373 women, with ages ranging between 18 and 88 years old (age $M$ = 46.47, s.d. = 15.09). Regarding race, 2291 participants reported being *White*, 181 *Asian*, 178 *Black*, 63 *Mixed* and 33 participants reported being from *other* ethnic groups. Additionally, two participants did not report their ethnicity. The majority of participants (94%) reported living in the United Kingdom (1817), United States (686) or Canada (72). Most of the participants (1482) reported having full-time jobs and 260 reported being students. More details about the participants can be found in appendix D.

Participants received a compensation of 2 USD for taking part in the study, with a median completion time of 8 min and 45 s. Seventeen participants failed at least two of the four attention checks and hence were removed from the analysis and replaced by new participants, yielding a total sample of 2748 participants.

## 4.2. Experimental stimuli

The stimuli used in the study were face images from two widely used face datasets for scientific research: the Chicago Face Database (CFD) [53] and the FACES dataset [54], and their corresponding beautified versions.

The CFD [53], developed at the University of Chicago for research purposes, provides high-resolution, standardized photographs of 597 unique individuals (male and female faces) of varying ethnicity (self-identified White, Asian, Black, Latino) between the ages of 17 and 56. The dataset was expanded in 2020 to include images of 88 mixed-race individuals recruited in the United States and 142 individuals recruited from India. While there are examples of faces with non-neutral facial expressions, we selected the images where all individuals have neutral facial expressions, yielding a dataset of 827 images. In addition to the images, the CFD dataset includes metadata about each image, such as information about physical attributes (e.g. face size) and subjective ratings by independent judges (e.g. attractiveness). The set of images collected from India has ratings available from both Indian and American raters. However, we used only the ratings of American raters in order to be consistent with the ratings for other images in the dataset. While the CFD includes a broad range of subject ages in their images, it mostly contains images of young people. Only 9% of the images are of people rated as being over 40 and it contained no images of people rated as being older than 60.[3] Thus, to ensure age diversity in the stimuli, we also included images from the FACES dataset [54].

The FACES dataset consists of 171 images of naturalistic faces of young ($n$ = 58), middle-aged ($n$ = 56) and older ($n$ = 57) women and men displaying each of six facial expressions: neutral, sadness, disgust, fear, anger and happiness. The database comprises two sets of pictures per person and per facial expression, resulting in a total of 2052 images. We selected the images corresponding to a neutral facial expression to minimize the interference of the facial expressions in the perception of attractiveness [54,111]. In addition to the images, the dataset includes metadata about each image, including subjective ratings of attractiveness from independent judges.

To ensure a balanced sample across age, gender and attractiveness levels, we selected 25 images[4] for each gender–ethnicity pair from the CFD, covering a wide spectrum of attractiveness levels: eight images with the lowest attractiveness ratings, eight images with the highest attractiveness levels and nine randomly selected from the remaining images. Similarly, we selected 27 images for each gender–age group pair from the FACES dataset ensuring diversity in gender, age and attractiveness levels. Since the FACES dataset had two images for each subject, we selected one at random. This process led to a total of 462 images (300 from the CFD and 162 from the FACES dataset) which we refer to as the **PRI** dataset of images (**P**icked **R**epresentative **I**mages). The subset that comes from the CFD is referred to as the $\texttt{PRI}_{\text{CFD}}$ dataset and similarly the images drawn from the FACES database are referred to as the $\texttt{PRI}_{\text{FACES}}$ dataset of images. A summary of these datasets can be found in table 3. Each face in the PRI dataset was *beautified* using a common beautification filter available in one of the most popular selfie editing apps in the world with over 500 million downloads. We refer to the dataset of beautified images as the **POST** dataset (**PO**st **S**ocial media **T**ransform). The filters were applied by running the selfie editing app on an Android emulator. An automated clicker loaded the pictures onto the application, applied the filter and then stored the transformed version.

[2]Link: https://doi.org/10.17605/OSF.IO/AQDK9.

[3]The CFD does not include the actual age of the participants in the pictures. Thus, the statistics about the age reported here are based on estimated age ratings from independent judges hired by Ma *et al*. while creating the CFD.

[4]The distribution of images across gender-ethnicity pairs in the CFD is non-uniform. The smallest class (Mixed Race Males) contained 26 images, thus motivating the size of the number of images picked for each gender–ethnicity pair.

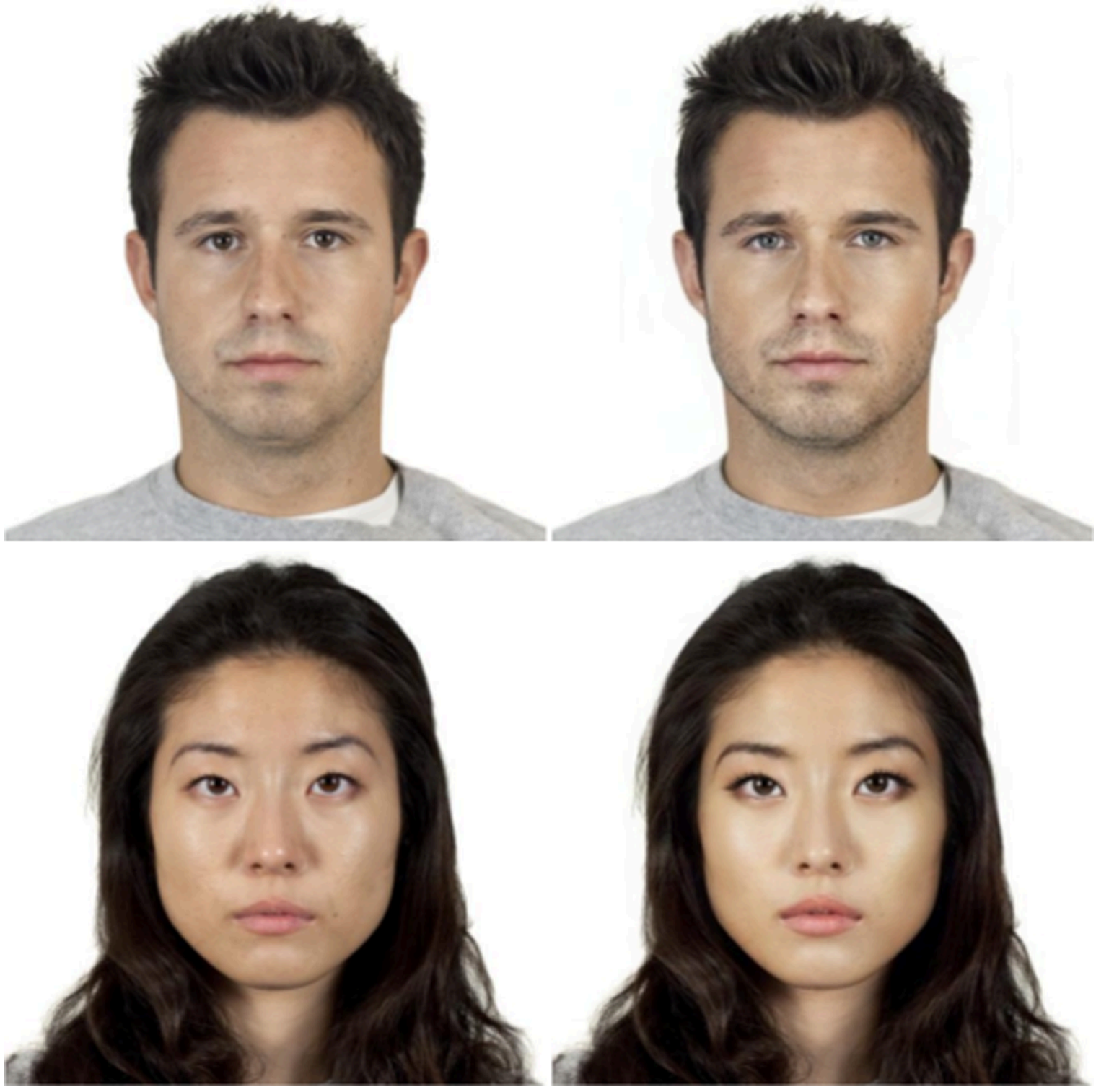

**Figure 6.** Samples of male (top) and female (bottom) face images used in our study before (left) and after (right) the application of the beauty filter. As illustrated in the examples, the beauty filter modifies the skin tone, the eyes and eyelashes, the nose, the chin, the cheekbones and the lips in order to make the person appear more attractive.

**Table 3.** Dataset statistics. Size corresponds to the number of unique faces present in the dataset. Age is the age of the subject in the image when the picture was taken (^ as perceived by the raters used by Ma *et al*. [53]. Actual age of the subjects in the images is not available.)

| | PRI | | | |
|---|---|---|---|---|
| | $\texttt{PRI}_{\text{CFD}}$ | $\texttt{PRI}_{\text{FACES}}$ | CFD | FACES |
| size | 300 | 162 | 827 | 171 |
| age | 18–56 ^ | 19–80 | 17–56 ^ | 19–80 |
| gender | 150M, 150 F | 81M, 81F | 406M, 421F | 86M, 85F |
| ethnicity | Asian, Black, Latino, White, Indian, Mixed | White | Asian, Black, Latino, White, Indian, Mixed | White |

Figure 6 shows an example of male and female original and beautified faces used in our study.

## 4.3. Procedure and design

The study was run online by means of a custom-made web portal. After providing informed consent, each participant was presented a page with instructions: they were told that they would be shown 10 face images and would be asked to provide their assessment of different aspects of the faces, based on their first impression. The exact instructions can be found in appendix E. Participants were randomly assigned to see faces either from the FACES dataset or the CFD.

After reading the instructions, participants were shown one face image at a time. Each image was accompanied with a set of questions as described in §4.4. The 7-point Likert rating scales were presented as sliders with the end- and mid-points labelled. Participants were required to answer all questions about an image before being allowed to proceed to the next image. The order of the questions was randomized for each participant as per the algorithm described below, but remained the same across all the images rated by the same participant. After providing ratings for 10 images, participants reached the last page of the survey where they were asked to provide details about their background including how often they used social media and beauty filters and their self-rated attractiveness. The complete list of questions is included in appendix E. After answering these questions, the study was complete and participants were directed to the Prolific platform where they were compensated for their time. The data collected in the study has been deposited in a public online repository [112].

In addition to the questions described in §4.4, participants were also shown four attentiveness checks (described in appendix E) at random points in the survey. Participants who failed two or more attention checks were rejected and additional participants were recruited to replace them.

The randomization algorithm to select the images that were shown to each participant met the following criteria to ensure a balanced sample:

(1) Half the images were from the `POST` dataset, i.e. had a beauty filter applied on them, and the other half were from the `PRI` dataset. The presentation order of the images was randomized and participants were not told that some of the images were beautified. Furthermore, participants always rated images corresponding to 10 different individuals such that they never had to judge the same person in both the beautified and non-beautified conditions.
(2) Half the images corresponded to male and the other half to female subjects.
(3) The images were also balanced across ethnicity (for participants in the CFD condition) or across age groups (for participants in the FACES condition).

Furthermore, the images were presented such that each image received at least 25 ratings.[5] Thus, participants provided ratings on a diverse set of inputs while ensuring that each image received sufficient ratings. Note that no image received ratings from the same subset of participants. Our analyses are adjusted accordingly.

## 4.4. Measures

For each image, participants were first asked to provide the gender (male/female), age (number between 18 and 100, answered using a sliding scale) and ethnicity (Asian/Black/Latino/White/Indian/Mixed Race) of each of the faces.

Next, participants were asked to rate the person in the image on the following attributes, which were randomly presented for each participant: physical attractiveness, intelligence, trustworthiness, sociability, happiness, femininity and how unusual they were. The choice of using intelligence, trustworthiness, sociability and happiness as dependent attributes for the halo effect was driven by existing literature on this cognitive bias, such as [4,6,8,9,22,53,60–63]. Ratings for femininity and unusualness were collected to study the impact of the beauty filters on physical appearance. Results of the analysis of the data corresponding to these two attributes have been discussed in detail in appendix C.

The ratings for attractiveness and other attributes were provided on a 7-point Likert scale ranging from 1 = *Not at all [trait term]* to 7 = *Extremely [trait term]*. While some work collected these ratings on a 9-point Likert scale [22], we opted to use a 7-point Likert scale because they have been reported to be the most accurate and reliable [113–115], despite the popularity of 5-point scales. Each question was presented to participants as 'How [trait term] is this person?', following the same approach as previous studies in the literature [6,53,61,106–109]. The responses were entered on a slider initially placed at the mid-point, and where both the mid and end points were labelled. An example of the layout of the questions participants were exposed to, along with the exact phrasing of the questions, can be found in appendix E.

[5]After rejections, two images were left with 23 ratings. The largest number of ratings received by an image was 35. The mean number of ratings each image received was 29.7 with a standard deviation of 1.75.

## 4.5. Analysis

As seen in figure 1, our analysis is structured according to two levels of aggregation: (i) *centralized scores*, by computing the median of the ratings provided to each image; and (ii) *individual scores*, by analysing the per-image ratings individually. As a result of our study methodology, each image received ratings from a different subset of participants such that pairwise comparisons are only possible at an aggregate level by means of the centralized scores. Furthermore, the change in the centralized scores ($\Delta_\omega$) is used as a measure of the impact of the filter. While analysing the impact that the age, gender and ethnicity of the stimuli play on perceptions on attractiveness and the four dependent variables by means of the centralized scores, we use the actual age, gender and ethnicity of the individuals in the image instead of the age, gender and ethnicity as perceived by the raters.

In order to study the effect of the participants' age and gender on attractiveness and the dependent variables, we also analyse each rating individually. All the variables collected in our study, except age, were collected on 7-point Likert scales, i.e. they are ordinal in nature. A multi-nomial logistic regression approach would treat the variables as nominal, thereby leading to a loss of information due to ignoring the inherent ordering of the responses. Using ordinal response models such as the cumulative link model (CLM) [116,117] is more appropriate for ordinal response variables [118] but the parameters of these models are harder to interpret than those of a linear model [119]. OSMs [57–59] offer an ideal middle ground. The OSM estimates the true spacing between the points on the ordinal scale based on the data, thereby resulting in a transformed scale that is continuous and thus suitable for a linear model. In addition to the theoretical grounding, we further evaluated the appropriateness of using the OSMs with linear models (and linear mixed models) by computing the Akaike information criterion (AIC) [120] and the Bayesian information criterion (BIC) [121] of different models for attractiveness and each of the four dependent variables. We also varied the treatment of the raters as fixed or random effects in our models. We found that linear mixed models with the OSMs that treat the raters as random effects resulted in the best (lowest) AIC and BIC scores. A detailed report of this analysis can be found in appendix J and the model parameters of the resultant linear mixed models can be found in appendix L.

Next, we describe in detail the OSMs (§4.5.1) and the linear mixed models (§4.5.2) that we developed to perform this second level of analysis. All the modelling has been performed in R version 4.3.3 [122].

### 4.5.1. Ordinal data

#### 4.5.1.1. Ordered stereotype models

Given an ordinal response variable $Y$ with $q$ categories, for an observation $i$, the OSM estimates the probability of $Y_i = k (k = 1 \cdots q)$ as

$$log\left(\frac{P[Y_i = k \mid x_i]}{P[Y_i = 1 \mid x_i]}\right) = \alpha_k + \phi_k \beta' x_i, \tag{4.1}$$

where $x_i$ is a set of predictor covariates for observation $i$. The OSM additionally enforces the constraint

$$0 = \phi_1 \leq \phi_2 \leq \cdots \leq \phi_q = 1\,. \tag{4.2}$$

The $\phi_k$'s are interpreted as scores and help estimate the distance between different categories based on the actual data (ratings in our case) instead of assuming that all categories are equidistant. Furthermore, categories with overlapping standard deviation intervals are merged into the same category. Thus, the OSMs estimate the underlying scale by computing the expected probabilities of the categories based on potential covariates in the data.

To evaluate the impact of the stimulus's and rater's gender and age on perceptions of attractiveness, the following OSM was fit independently to the data from the `PRI` and `POST` sets:

$$Attrac \sim Gender_I + Age_I + Gender_R + Age_R, \tag{4.3}$$

where $Gender_R$ and $Age_R$ correspond to the gender and age of rater $R$, respectively, and $Gender_I$ and $Age_I$ correspond to the gender and age of image $I$ as perceived by rater $R$. In the case of the dependent variables ($\omega$), perceived attractiveness was also included as a covariate,

$$\omega \sim Attrac_I + Gender_I + Age_I + Gender_R + Age_R\,. \tag{4.4}$$

### 4.5.2. Linear mixed models

Below are the linear mixed models presented in §2.3. Note that linear mixed models on the rescaled data by means of the OSM are a better fit to the data than ordinal models, such as CLMs [116], as explained in appendix J. Furthermore, linear mixed models that include the raters as random effects better fit the rescaled data than models that treated the raters as fixed effects (see appendix J),

$$\begin{aligned} Attrac = \beta_0 + \beta_2 \cdot Gender_I + \beta_3 \cdot Age_I + \beta_4 \cdot Gender_R + \\ \beta_5 \cdot Age_R + \beta_6 \cdot Gender_I \cdot Gender_R + \beta_7 \cdot Age_I \cdot Age_R + RandEff_{Rater}, \end{aligned} \tag{4.5}$$

$$\begin{aligned} \omega = \beta_0 + \beta_1 \cdot Attrac_I + \beta_2 \cdot Gender_I + \beta_3 \cdot Age_I + \beta_4 \cdot Gender_R + \\ \beta_5 \cdot Age_R + \beta_6 \cdot Gender_I \cdot Gender_R + \beta_7 \cdot Age_I \cdot Age_R + RandEff_{Rater}\,. \end{aligned} \tag{4.6}$$

The above models consider the stimulus's age ($Age_I$) and gender ($Gender_I$) as perceived by the rater, the rater's self-reported gender ($Gender_R$) and age ($Age_R$) and the interactions between these variables. Race was not included as a variable in the analysis because the previously reported results with the centralized ratings revealed no significant impact of race, neither on attractiveness nor on the dependent attributes. Additionally, the participants' self-reported race was predominantly white (see appendix D) and hence it was also not considered as a variable in the models. Note that $\beta_1$ is omitted from the linear mixed model of attractiveness (4.5) to maintain consistency in the terminology, since the linear mixed models of the dependent variables (4.6) use $\beta_1$ for attractiveness. The models were fit independently on the `PRI` and `POST` sets. The parameters of all the linear mixed models can be found in appendix L.

**Ethics.** The experimental procedures in this study were approved by the ethics committee of the University of Alicante with identifier UA-2023-01-19_3. All participants gave informed consent.

**Data accessibility.** The data collected by us during the survey and the associated code can be found in the following Zenodo repository: [123]. The code, along with Zenodo repository, can also be found in the following GitHub repository: [124]. The images from the PRI set are publicly available and access instructions can be found in the Readme file on Zenodo and GitHub. The POST set images used in this study were created using a common beautification filter available in one of the most popular selfie editing apps as indicated in the manuscript. Our agreement with the application provider does not allow publicly sharing the beautified images. For further queries about this data, please contact the legal department of ELLIS Alicante at info@ellisalicante.org.

**Declaration of AI use.** We have not used AI-assisted technologies in creating this article.

**Authors' contributions.** A.G.: conceptualization, data curation, formal analysis, investigation, methodology, project administration, resources, software, validation, visualization, writing—original draft, writing—review and editing; M.M.-G.: formal analysis, investigation, methodology, supervision, validation, visualization, writing—review and editing; D.F.: formal analysis; M.A.L.: methodology, project administration; B.L.: conceptualization, formal analysis, investigation, methodology, supervision, writing—review and editing; N.O.: conceptualization, data curation, formal analysis, funding acquisition, investigation, methodology, project administration, resources, supervision, validation, visualization, writing—original draft, writing—review and editing.

All authors gave final approval for publication and agreed to be held accountable for the work performed therein.

**Conflict of interest declaration.** We declare we have no competing interests.

**Funding.** A.G. and N.O. are partially supported by a nominal grant received at the ELLIS Unit Alicante Foundation from the Regional Government of Valencia in Spain (Convenio Singular signed with Generalitat Valenciana, Conselleria de Innovacion, Industria, Comercio y Turismo, Direccion General de Innovacion), along with grants from the European Union's Horizon 2020 research and innovation programme (ELIAS; grant agreement 101120237) and Intel. A.G. is additionally partially supported by grants from the Banc Sabadell Foundation and the European Union's Horizon 2020 research and innovation programme (ELISE; grant agreement 951847). B.L. is partially supported by the European Union's Horizon Europe research and innovation programme under grant agreement no. 101120237 (ELIAS) and by the PNRR project FAIR—Future AI Research (PE00000013), under the NRRP MUR programme funded by the NextGenerationEU. M.M.-G. is supported by grant CIGE/2022/066 by Generalitat

Valenciana and grant PID2020-118071GB-I00 from the Ministerio de educación y formación profesional. D.F. is supported by the Ministerio de Ciencia e Innovación y Universidades (Spain) [PID2023-148033OB-C21], and by grant 2021 SGR 01421 (GRBIO) administrated by the Departament de Recerca i Universitats de la Generalitat de Catalunya (Spain).

# Appendix A. Impact of self-perceived attractiveness on judgements of attractiveness

An additional factor that might mediate the strength of the attractiveness halo effect is the perceived attractiveness of the human evaluators. Humans are unable to make self–target comparisons without assessing their own physical attractiveness [125]. Previous work has found that the self-concept of physical attractiveness is also associated with positive affect, cognitive and social measures [126]. Not only our behaviour is shaped by the levels of attractiveness that we perceive ourselves with, but we use such a self-assessment as a benchmark when determining the attractiveness of others. According to the in-group/out-group theory [127], in-group members share similar attributes and assign more positive attributes to each other than to out-group individuals [128]. Consequently, the self-perceived attractiveness of the human evaluators would impact their perception of the stimuli, leading to different perceptions of social distance between the self and the target. Recent work by Li *et al*. [129] has studied and corroborated this phenomenon in the context of consumer evaluation processes during a service encounter: the evaluators' (consumers in this case) perception of their physical attractiveness was found to play a moderating role on the attractiveness halo effect. However, other authors have not identified any interactions between self-perceived attractiveness and the halo effect in certain contexts, such as hireability [130].

In this section, we study the relationship between the participants' self-perceived attractiveness and their attractiveness judgements. Figure 7 depicts the histograms of (*a*) the self-reported attractiveness levels of the participants in our study (mean: 4.17, s.d.: 1.21), and (*b*) the centralized attractiveness scores reported by the same raters to images from the `PRI` set (mean: 3.57, s.d.: 0.91).

Given the self-perceived attractiveness of rater $R$, $R_{SRA}$; the attractiveness score provided by rater $R$ to stimulus $I$, $I^R$; and the stimulus centralized attractiveness score, i.e. its median attractiveness, $I^c$, we identify a weak correlation ($\tau = 0.049$, $p = 0.007$, Kendall) between the participant's self-perceived attractiveness and their attractiveness ratings overall, and both in the images belonging to the `PRI` ($\tau = 0.019$, $p = 0.008$, Kendall) and `POST` ($\tau = 0.053$, $p < 0.001$, Kendall) sets. Thus, self-perceived attractiveness has not been included as a factor in the reported analyses. While there is no strong relationship between the rater's self-perceived attractiveness and the attractiveness ratings they provide, an interesting phenomena does emerge in figure 7: more than 50% of the participants consider themselves to be above average on attractiveness (above a 4 on the 7-point Likert scale). We observe a clear distribution shift between the distribution of attractiveness scores provided to the stimuli (figure 7*b* and the distribution of self-rated attractiveness (figure 7*a*)). This finding is in line with literature that finds that humans tend to overestimate their qualities and abilities and underestimate those in others [131,132].

# Appendix B. Ordered stereotype models

Figure 8 shows the new scales computed using the OSM for attractiveness and each of the dependent attributes independently on the `PRI` and the `POST` datasets. Since the scales were computed independently on the `PRI` and `POST` datasets, each attribute has a different scale with possibly a different number of points in each set, making it harder to directly compare the ratings an image receives before and after the filters are applied.

The scale for attractiveness is compressed to six points in the `PRI` dataset and five points in the `POST` dataset. The highest/lowest values of the scale are merged in the `PRI` and `POST` datasets, respectively, because there are few ratings corresponding to such values. The scale for intelligence is compressed to a 5-point scale in the `PRI` dataset and a 6-point scale in the `POST` dataset. In both cases, the highest values of the scale are merged. The scale for trustworthiness is compressed to a 5-point scale both in the `PRI` and `POST` datasets. The highest values of the scale are merged in the `PRI` dataset and the levels corresponding to the [5,6] values in the original scale are merged in the `POST` dataset. The scale for sociability exhibits a similar behaviour to that of *attractiveness*, with a compression to a 6-point instead

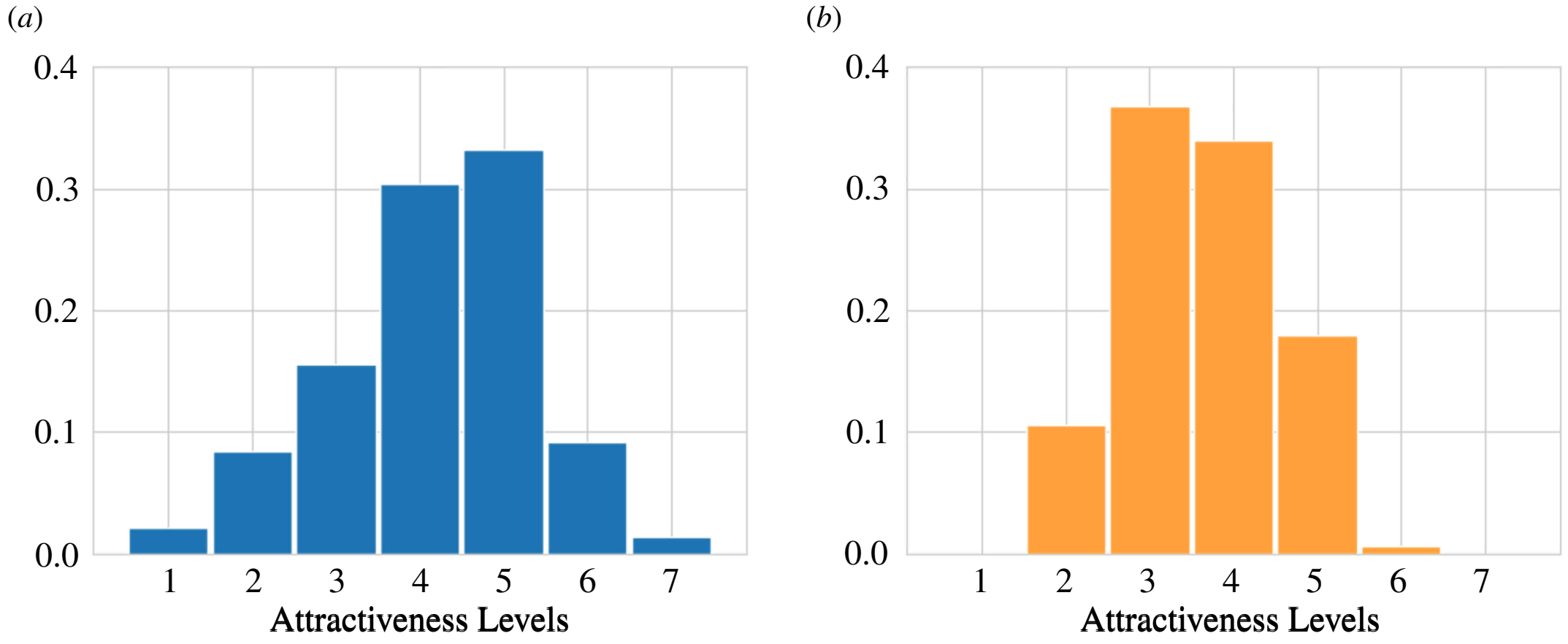

Distribution of the self-rated attractiveness of participants in our study

Distribution of the attractiveness ratings provided to images in the PRI set of a 5-point scale in the POST dataset. Interestingly, *happiness* is the only dependent attribute for which the scale remains as a 7-point scale both in the PRI and POST datasets, with an adjustment of the distance between consecutive points in the scale.

**Figure 7.** Distribution of attractiveness ratings of (*a*) the self-rated attractiveness of the participants and (*b*) the attractiveness ratings provided to images in the PRI set.

# Appendix C. Impact of beauty filters on physical attributes

Section 2.1 established that the filters manipulate the perceptions of attractiveness. However, it is not clear how the filters impact the perception of other physical characteristics such as age, gender and ethnicity. Section 2.1.1 below analyses this impact by performing pairwise tests. In addition, we explore how filters affect the characteristics related to physical appearance such as perceived femininity and unusualness in §2.1.2.

While we see that beauty filters have a slight but statistically significant impact on perceptions of age, gender and ethnicity, these factors play a small role in determining the perceptions of the dependent variables according to our linear mixed models. Appendix M presents the partial $R^2$ of all the predictor variables which indicate that attractiveness is the strongest predictor of each of the dependent attributes.

## C.1. Impact of beauty filters on perceptions of age, gender and ethnicity

Computing a centralized score enables a pairwise comparison between the perceptions of age, gender and ethnicity of the images before and after beautification. Perceptions of attractiveness and the dependent variables ($\omega$) were reported on an ordinal scale and perceptions of age were reported on a continuous scale, making the median a representative central tendency. Raters, however, were required to provide a categorical response when reporting the gender of the person they see in the image. Below is a summary of our findings.

### C.1.1. Age

We identify a statistically significant difference ($p < 0.001$, Wilcoxon paired-rank) in the centralized perceived age between images in the PRI and POST sets. The difference was also significant ($p < 0.001$) for all age groups, both genders and all ethnicities. The mean of the difference in perceived age ($\Delta_{Age} = Age_{\mathrm{POST}} - Age_{\mathrm{PRI}}$) was 5.87 years, indicating that filters reduce the perceived age of subjects in images significantly. The change in perceived age, however,was not the same for all images.

The reduction in perceived age after the application of the filters for different age groups is statistically significant. The mean change in age for images of young individuals (−2.32) was significantly less ($p < 0.001$, Kruskal–Wallis) than the change for middle-aged (−10.99) and older individuals (−7.69). The difference between middle-aged and older individuals was, however, less significant ($p < 0.01$, Kruskal–Wallis). There was also a significant difference ($p < 0.001$, Kruskal–Wallis) in the change in perceived age for images of females ($\Delta_{Age}= -6.86$) when compared with images of males ($\Delta_{Age}= -4.89$), but no differences across different ethnic groups.

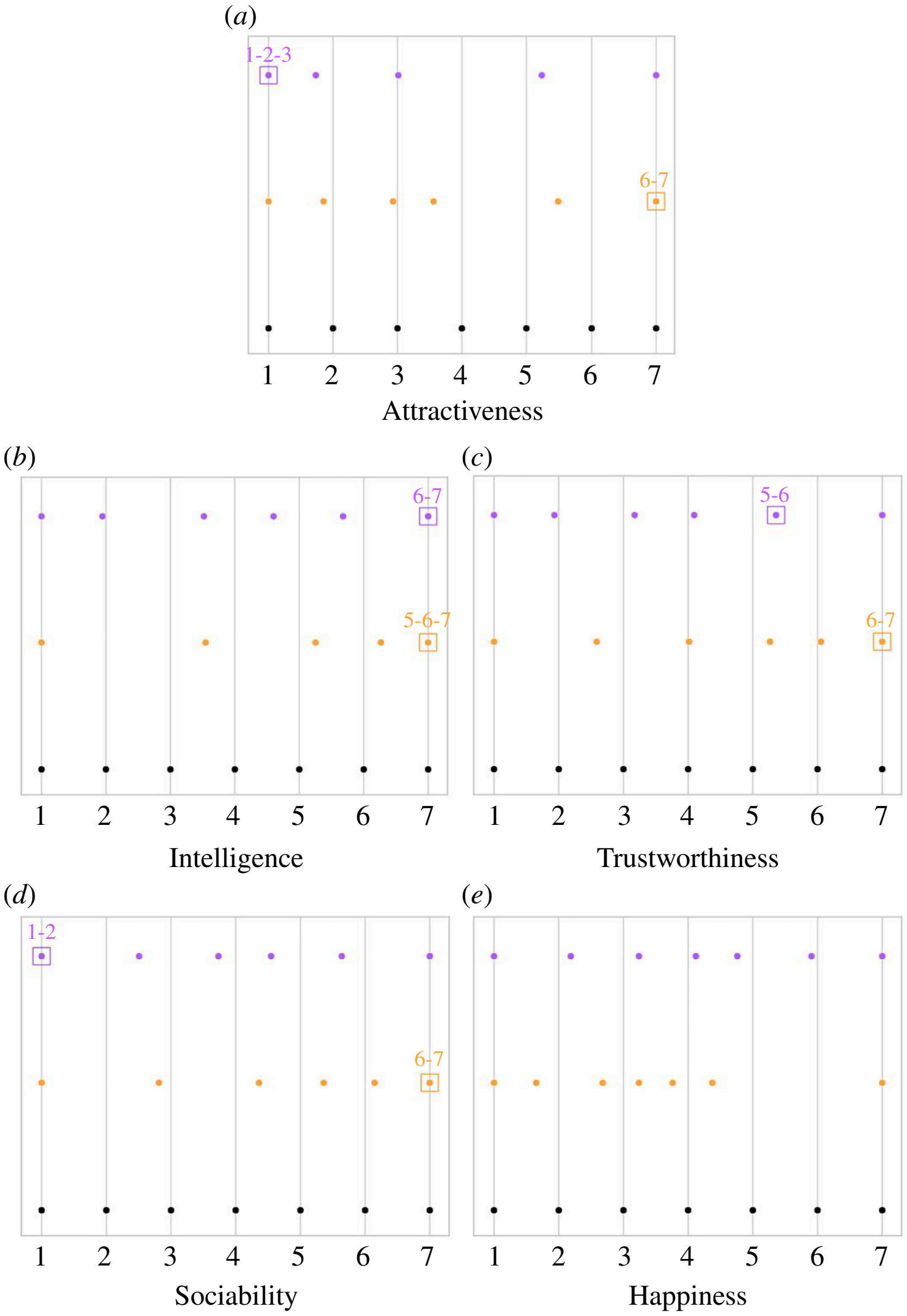

**Figure 8.** The new scales for perceived attractiveness and the dependent variables after rescaling by means of OSMs based on the collected data. The black dots at the bottom indicate the original, equally spaced 7-point Likert scale. The orange and purple dots correspond to the new scale for the PRI (.) and POST (.) sets, respectively. The squares around the dots indicate locations where multiple points on the scale were collapsed to the same value.

### C.1.2. Gender

In the case of gender, we compute the mis-classification rate, i.e. the fraction of participants whose predicted gender did not match the ground-truth gender of the individual in the image as provided by the image datasets. Our analyses revealed statistically significant ($p < 0.001$, Wilcoxon paired-rank) differences between the gender mis-classification percentage of the images in the PRI and the POST sets. The mis-classification rate was on average 0.006 lower in the POST set than the PRI set. This difference is more pronounced for images of females where the mean difference in the mis-classification rate was on average 0.01 points lower. Interestingly, for images of males, the differences were not significant.

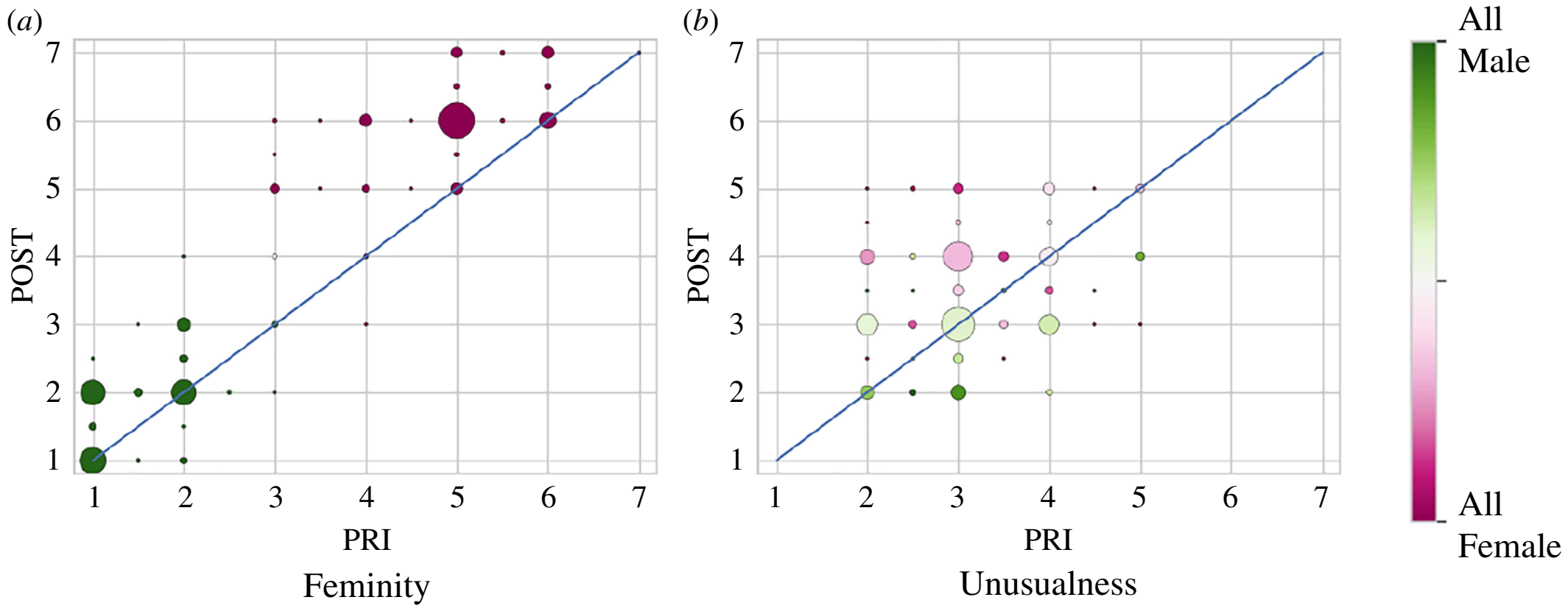

**Figure 9.** Comparison of the perceptions of (*a*) femininity and (*b*) unusualness before (*x*-axis) and after (*y*-axis) the application of beauty filters.

### C.1.3. Ethnicity

Similar to gender, reporting of ethnicity was also categorical. Thus, we use the mis-classification rate as a representative statistic. Statistically significant ($p < 0.001$, Wilcoxon paired-rank) differences were found between the mis-classification rate of the ethnicity of images in the PRI and the POST sets. The mis-classification rate was on average 0.042 lower in the POST set when compared with the PRI set. Thus, the filters do impact the perception of ethnicity of subjects.

## C.2. Impact of beauty filters on perceptions of femininity and unusualness

Beauty filters have been hypothesized to project female faces closer to normative ideals of femininity [133] and have been shown to homogenize faces [39,47]. Thus, participants were asked to rate images on perceived femininity and perceived unusualness, i.e. would they stand out in a crowd.

Perceptions of femininity and unusualness increased significantly ($p < 0.001$, one-sided Wilcoxon paired-rank) after the application of beauty filters. The impact of the filters on perceived femininity, measured through the $\Delta_{Femininity}$, was significantly ($p < 0.001$, Kruskal–Wallis) higher for images of females (mean increase of 0.98) than it was for images of males (mean increase of 0.35). This finding is illustrated in figure 9*a*, which depicts the perceived femininity scores before and after the filters were applied. Images of females received significantly higher femininity scores than images of males, as expected.

Similarly, the impact of the filters on perceived unusualness, measured through the $\Delta_{Unusualness}$, was significantly larger ($p < 0.001$, Kruskal–Wallis) for images of females (mean increase of 0.53) than it was for images of males (mean increase of 0.09). figure 9*b* shows the comparison of perceived unusualness scores of images before and after beautification. While images of females tend to exhibit a large increase in perceived unusualness, some images of males also exhibit a decrease in perceived unusualness. Thus, further studies are needed to understand the homogenizing effect of the filters.

# Appendix D. Participant information

Figure 10 summarizes the characteristics of all the participants included in our study as per their self-reported answers (see appendix E for the complete list of questions) and the information that they shared with Prolific as a part of their participant profile.

Regarding age, the age of the participants ranged between 18 and 88 years old (means: (33.22,57.99), covariances: (59.34,88.35)). In terms of social media usage, the majority (58.19%) of participants reported using social media several times a day (figure 10*b*), Facebook (35.45%) and Instagram (27.80%) being the most used social platforms, as depicted in figure 10*c*. Most participants (79.66%) responded to never use beauty filters while posting content on social media (figure 10*d*).

Figure 10*e* depicts the Kendall correlations between these characteristics. The strongest positive correlation (0.36) is found between the usage of filters and the posting frequency on social networks,

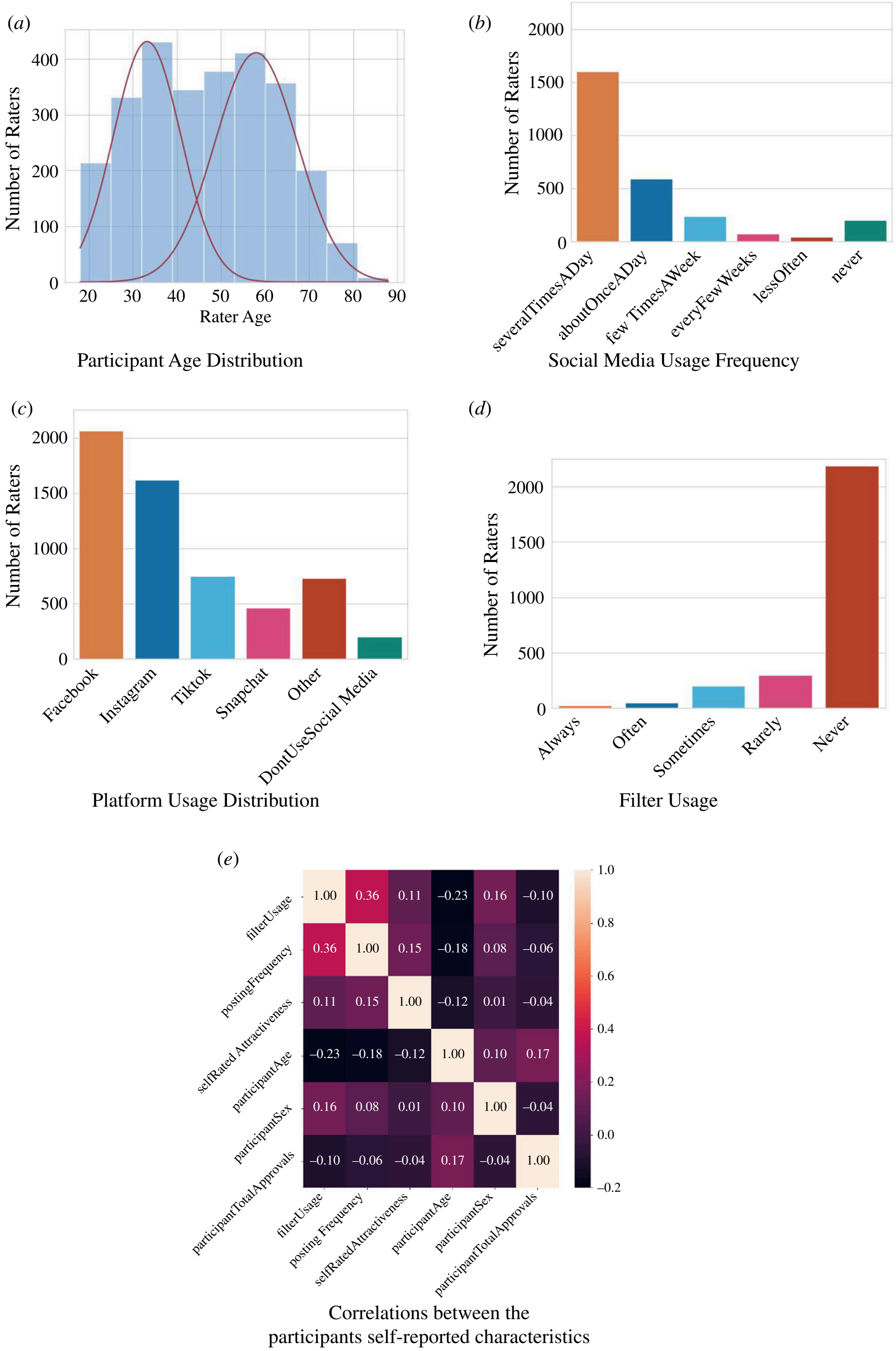

**Figure 10.** A summary of the characteristics of the 2748 participants of our study.

whereas the strongest negative correlation (−0.23) is identified between filter usage and the age of the participant. Interestingly, we observe also slight positive correlations between filter usage and the participant's sex (0.16) and between self-perceived attractiveness and the posting frequency (0.15); and a slight negative correlation (−0.12) between the participant's age and their self-perceived attractiveness. As there was no significant correlation between any of these variables and the attractiveness ratings provided by participants, they were excluded from our analysis.

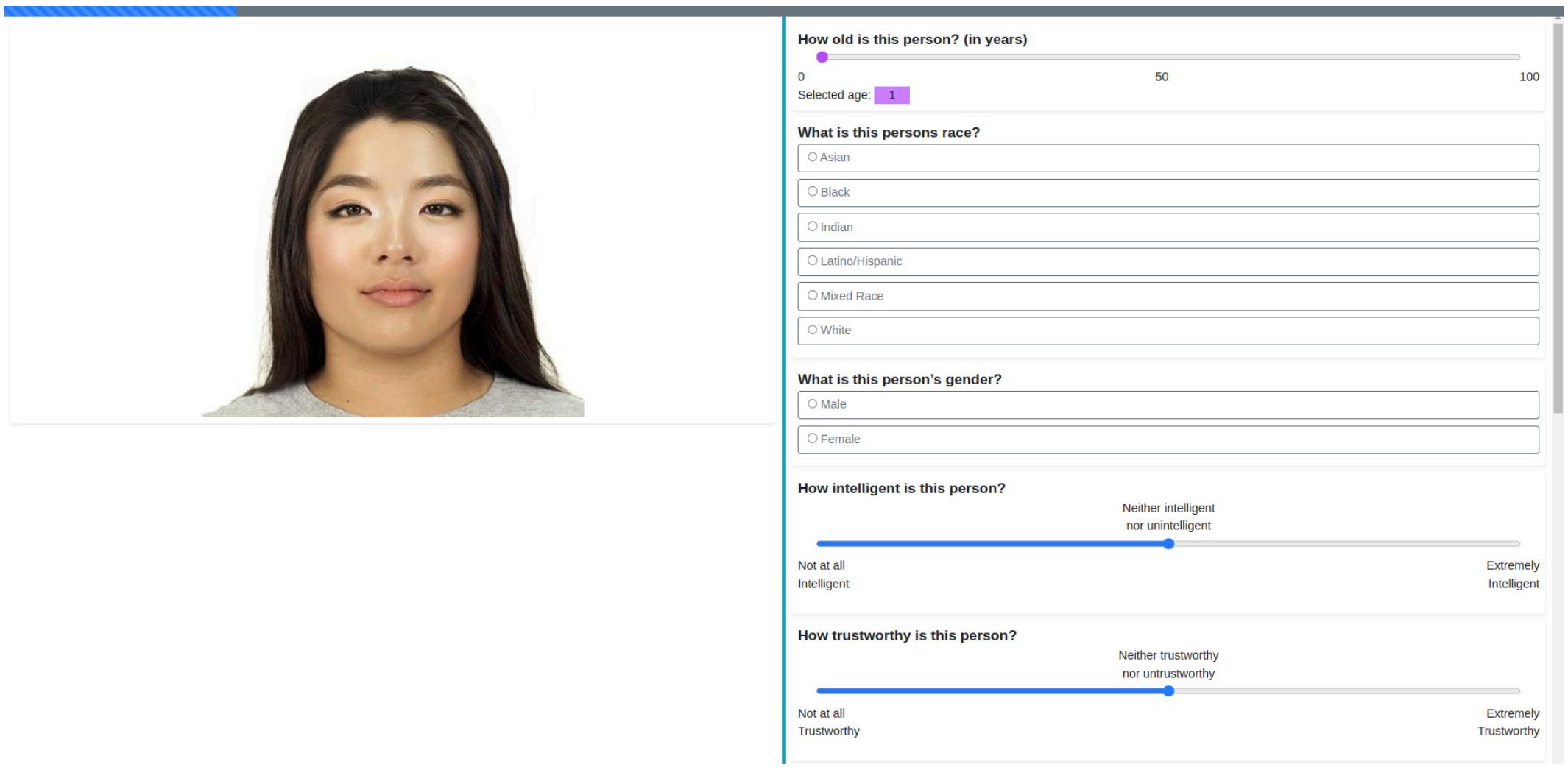

**Figure 11.** A screen shot of the survey tool. The face image on the left remains static on the screen as participants scroll through the column on the right to answer all the questions described in §4.4 about each image. A progress bar on the top indicates progress in answering the questions.

# Appendix E. Survey design

## E.1. The survey tool

The survey was administered through a custom-made web portal created by the first author and illustrated in figure 11. Participants who signed up for the study accessed the survey through their web browsers and were encouraged to use the tool from a laptop/desktop, to ensure a similar user experience among participants. The website did not use cookies and the participants' responses were tracked through an anonymized identifier generated by Prolific which was shared with the tool when the survey was started. After receiving the instructions described in appendix E.2, participants saw a face image on the left half of the screen and a set of questions corresponding to the image on the right half. The image was fixed on the screen such that participants were able to scroll through the questions while having access to the image. Participants were required to provide answers for every question before moving to the next face image. The survey tool was optimized to prevent data loss in between responses and to ensure a smooth user experience. All data were stored on the user's web browser until it was sent to a secure AWS database.

## E.2. Instructions

Upon entering the survey, participants were first asked to confirm that they were adults (18+ years old) and to consent to participating in the study. Only those who responded affirmatively to both questions proceeded to a new page with the instructions below:

- This study consists of two parts. In the first part, you will see a person's face, and will be asked to judge them on a small collection of attributes. Importantly, go with your **gut feeling**. We all make snap judgements of others constantly, so feel free to report what you think about the person based on their face. Please respond quickly with your gut feeling. **There are no right or wrong answers.**
- The faces you see might have different ethnicities. When you provide your ratings for a person, please try to rate them with respect to other people of the same race and gender. (For example, if you indicated that the person was Asian and male, consider this person relative to other Asian males.)
- Once you provide ratings on all the faces that have been randomly assigned to you, you will see a short questionnaire with a few questions about you. We will not ask for any personal identifying information.
- After you answer all the questions in both parts, you will automatically be redirected to Prolific's website and will receive your compensation for participating in our study.

## E.3. Questions

Every participant was sequentially shown 10 images corresponding to 10 distinct individuals. For each image, participants responded to the questions below. Participants were required to answer all questions before being allowed to proceed to the next image and were not allowed to revisit an image once they had provided their answers to all the questions corresponding to the image. The order of the questions was randomized for each participant, but stayed the same across all the images that they saw.

Q1. How old is this person? (in years)?

Participants were asked to respond on a 0 to 100 scale starting at 0. If participants entered an age less than 18, they were shown an error message below the age question which said 'Age needs to be between 18 and 100'.

Q2. What is this persons race?

- ○ Asian
- ○ Black
- ○ Indian
- ○ Latino/Hispanic
- ○ Mixed Race
- ○ White

Q3. What is this person's gender?

- ○ Male
- ○ Female

For the remaining questions, participants were asked to provide their answers on a 7-point Likert scale presented as a slider (indicated with a → symbol below) where the end and middle points are labelled, as per previous research [6,53,61,106–109].

Q4. How attractive is this person?

→ Not at all Attractive ... Neither attractive nor unattractive ... Extremely Attractive

Q5. How feminine is this person?

→ Not at all Feminine ... Neither feminine nor masculine ... Extremely Feminine

Q6. How unusual is this person? (Would they stand out in a crowd)

→ Not at all Unusual ... Neither unusual nor usual ... Extremely Unusual

Q7. How trustworthy is this person?

→ Not at all Trustworthy ... Neither trustworthy nor untrustworthy ... Extremely Trustworthy

Q8. How sociable is this person?

→ Not at all Sociable ... Neither sociable nor unsociable ... Extremely Sociable

Q9. How intelligent is this person?

→ Not at all Intelligent ... Neither intelligent nor unintelligent ... Extremely Intelligent

Q10. How happy is this person?

→ Not at all Happy ... Neither happy nor unhappy ... Extremely Happy

### E.3.1. Background information

After answering the above questions for 10 face images, participants were ask to respond to five questions (BQ1 to BQ5 below) related to their social media usage and their self-perception of attractiveness. BQ1 is a multiple choice question (indicated with a □ symbol below), BQ2–BQ4 are single choice (indicated with a ∘ symbol below) and BQ5 is a 7-point Likert scale question on a slider (→).

BQ1. Which of the following social media platforms do you use?

☐ Instagram
☐ Facebook
☐ TikTok
☐ Snapchat
☐ Other
☐ I do not use any social media platforms

BQ2. How often do you check into your social media accounts?

- Several times a day
- About once a day
- A few times a week
- Every few weeks
- Less often
- Never

BQ3. How often do you post pictures of yourself on social media?

- Several times a day
- About once a day
- A few times a week
- Every few weeks
- Less often
- Never

BQ4. When you upload an image of yourself on social media, do you apply beauty filters on the image?

- Always
- Often
- Sometimes
- Rarely
- Never

BQ5. How attractive would you say you are?

→ Not at all Attractive ... Neither attractive nor unattractive ... Extremely Attractive

## E.4. Attentiveness checks

Participants were also shown four attentiveness questions randomly placed in the survey. The appearance of these questions (sliders and options) was identical to the other questions presented in the survey. These attentiveness checks were compliant with Prolific's *attentiveness check policy*.[6] They were evaluated and approved by Prolific before deploying the survey.

The attentiveness checks shown to participants were randomly selected from the following pool of six questions:

We would like to ensure only real people answer our survey. To show that you are human, please move the slider below to 'Strongly Disagree'.

→ Strongly Disagree ... Neither disagree nor agree ... Strongly agree

[6] https://researcher-help.prolific.com/en/article/fb63bb

We would like to ensure only real people answer our survey. To show that you are human, please move the slider below to 'Strongly Agree'.

→ Strongly Disagree ... Neither disagree nor agree ... Strongly agree

We would like to ensure only real people answer our survey. To show that you are human, please click the button that says 'False' below.

- True
- False

We would like to ensure only real people answer our survey. To show that you are human, please click the button that says 'True' below.

- True
- False

We would like to ensure only real people answer our survey. To show that you are human, please click the button that says 'Blue' below.

- White
- Black
- Blue
- Green
- Yellow

We would like to ensure only real people answer our survey. To show that you are human, please click the button that says 'Tuesday' below.

- Monday
- Tuesday
- Wednesday
- Thursday
- Friday

# Appendix F. Factor analysis

Principal component analysis (PCA) of the centralized ratings in the `PRI` and `POST` datasets separately was performed for each dependent variable to identify correlations between them. Figure 12 depicts the projections of the data for each dependent variable on a two-dimensional space of the directions of the eigenvectors with the largest eigenvalues.

While sociability and happiness appear to be closely related in the `PRI` dataset, all four dependent attribute vectors are clearly separated in the `POST` dataset, with intelligence and sociability being almost orthogonal to each other. Based on these results, we perform analyses on these four dependent attributes.

# Appendix G. Impact of filters on dependent attributes

The images after beautification (`POST`) were rated significantly higher ($p < 0.001$, one-sided Wilcoxon paired-rank) on all dependent attributes when compared with their original (`PRI`) counterparts, as discussed in §2 and appendix I. Kruskal–Wallis ($\chi^2$) tests on the centralized scores, followed by pairwise Wilcoxon tests for each dependent variable ($\omega$)—namely intelligence, trustworthiness, sociability and happiness—on the original (PRI) and beautified (POST) faces depending on the age, gender and ethnicity of the stimulus revealed some statistically significant differences according to age and gender, but no statistically significant differences according to ethnicity, as was also observed with the perceptions of attractiveness.

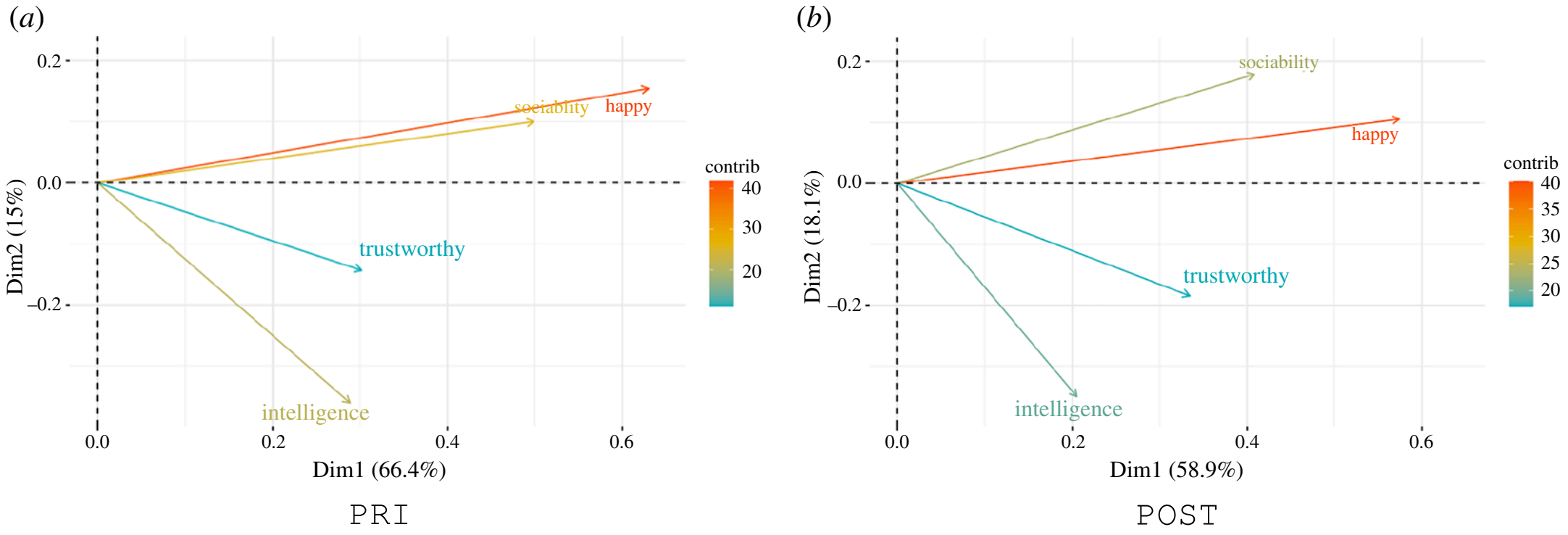

**Figure 12.** Projections of all the dependent attributes on the first two dimensions.

Younger individuals were perceived to be significantly ($p < 0.001$, pairwise Wilcoxon) more sociable than middle-aged individuals in both the PRI and POST datasets. While younger individuals were also perceived as being significantly ($p < 0.001$, pairwise Wilcoxon) more sociable than older subjects in the POST dataset, the difference was less significant ($p < 0.01$, pairwise Wilcoxon) than in the PRI dataset. There were no statistically significant differences in the perception of sociability between middle-aged and older individuals in either set. Hess *et al.* [85] reported a decrease in perceived sociability for stimuli of elderly individuals. While the stimuli they used had only young and old individuals, our study also included images of middle-aged individuals. While studying the impact of age on sociability was not the primary goal of our study, our findings suggest that the decreased perception of sociability is not true only for the elderly, but could potentially impact even middle-aged individuals.

While none of the dependent attributes other than sociability showed significant ($p < 0.001$, Kruskal–Wallis) differences across age groups, the beauty filters impacted the change in perceived intelligence ($\Delta_{intelligence} = \text{Intelligence}_{\text{POST}} - \text{Intelligence}_{\text{PRI}}$) differently across different age groups. The increase in perceived intelligence ($\Delta_{intelligence}$) was significantly lower ($p < 0.001$, pairwise Wilcoxon) for younger subjects when compared with middle- and older-aged individuals. There was no significant difference in $\Delta_{intelligence}$ between middle- and older-aged subjects. The differences across age groups in the change of the centralized ratings for all the other attributes (i.e. $\Delta_{\omega}$) was not statistically significant.

The impact of gender on perceptions of the dependent attributes was more pronounced. In the POST dataset, images of females received significantly ($p < 0.001$, Kruskal–Wallis) higher ratings on all dependent attributes except intelligence, yet there was no statistically significant difference in the ratings provided to images of males. In the PRI dataset, women were perceived as significantly ($p < 0.001$, Kruskal–Wallis) more trustworthy and sociable. Thus, we conclude that the filters enhanced the differences in perception of the dependent attributes between men and women. The change in the perception of the dependent attributes ($\Delta_{\omega}$) due to the filters was also different across genders. Images of females experienced a significantly larger increase ($p < 0.001$, Kruskal–Wallis) in perceptions of happiness ($\Delta_{happiness}$) and a less significant increase ($p < 0.01$, Kruskal–Wallis) in perceptions of sociability ($\Delta_{sociability}$). Interestingly, $\Delta_{intelligence}$ was also slightly significantly different ($p < 0.01$, Kruskal–Wallis) for images of males versus females, even though there was no significant difference in the perceived intelligence of the images depicting males versus females in either the PRI or POST datasets: images of males increased the scores in perceived intelligence more than images of females due to the filters. These findings are summarized in table 4.

While we identified a slightly significant ($p < 0.01$, Kruskal–Wallis) impact of ethnicity on perceptions of intelligence in the POST dataset, pairwise Wilcoxon tests did not reveal any statistically significant differences. Thus ethnicity, similarly as in the case of attractiveness, does not seem to impact the perceptions of the dependent attributes.

**Table 4.** Kruskal–Wallis ($\chi^2$) test on the median perceived values of each dependent variable in the original (PRI) and beautified (POST) faces depending on the age, gender and ethnicity of the individual. *** denotes $p < 0.001$; ** denotes $p < 0.01$ and * denotes $p < 0.05$.

| dependent variable ($\omega$) | image age (FACES) | | image gender | | image ethnicity (CFD) | |
|---|---|---|---|---|---|---|
| | PRI | POST | PRI | POST | PRI | POST |
| intelligence | 11.59** | 6.18* | 4.19* | 1.46 | 13.05* | 15.53** |
| trustworthiness | 4.02 | 10.78** | **26.41*** ** | **28.16*** ** | 5.37 | 7.41 |
| sociability | 13.73** | **18.76*** ** | **16.11*** ** | **62.68*** ** | 7.80 | 9.56 |
| happiness | 6.56* | 1.91 | 3.35 | **40.45*** ** | 11.25* | 3.17 |

**Table 5.** Differences (expressed as percentage of change) in the gender gap between the ratings provided to images of male and female subjects, by male and female raters, between the POST and the PRI datasets. There is a larger increase in the gender gap for male raters when rating attractiveness, intelligence and trustworthiness. However, there is a larger increase of the gender gap in the ratings provided by female raters when rating sociability and happiness. Statistical significance of the differences in the PRI and POST datasets are described in figure 4.

| dependent attribute ($\omega$) | male raters | female raters |
|---|---|---|
| attractiveness | **85.12** | 17.83 |
| intelligence | **331.23** | 158.81 |
| trustworthiness | **−95.68** | −37.74 |
| sociability | 49.60 | **1598.85** |
| happiness | 114.00 | **212.69** |

# Appendix H. Impact of gender of the rater

Figure 4 shows the expected marginal means (EMMs) of the ratings for images of males and females by male and female raters. For most attributes, both male and female raters provide different ratings to images of males and females. We refer to this difference in rating between images of males and females as a gender gap. This gender gap in ratings is impacted by the filters and depends on the gender of the rater.

Table 5 shows the differences (in percentage) of the gender gap between the ratings provided to the images in the PRI and the POST datasets by male and female raters for attractiveness and all the dependent attributes. Regarding perceptions of attractiveness, intelligence and trustworthiness, observe how male raters are more impacted by the filters as there is a larger gender gap when compared with female raters. However, for perceptions of sociability and happiness, female raters are more impacted by the filters than male raters.

# Appendix I. Wilcoxon paired-rank tests

Table 6 summarizes the results of the Wilcoxon paired-rank tests on the centralized scores of attractiveness and the dependent variables. The test statistic is positive and significant ($p < 0.001$), indicating that the beauty filters significantly impacted perceptions of attractiveness and all the dependent variables. Additionally, figure 13 depicts a visualization of the change in centralized scores after applying the beauty filter. The $x$-axis and $y$-axis correspond to the 7-point Likert scale scores of the images in the PRI and POST sets, respectively. The size of the circles is proportional to the number of images with the corresponding values. The colour of the circles reflects the proportion of male/female faces represented by that point. While there were no images with a decrease in their attractiveness score after applying the beauty filters (figure 2*a*), observe there were images with lower scores in intelligence, trustworthiness, sociability or happiness after beautification.

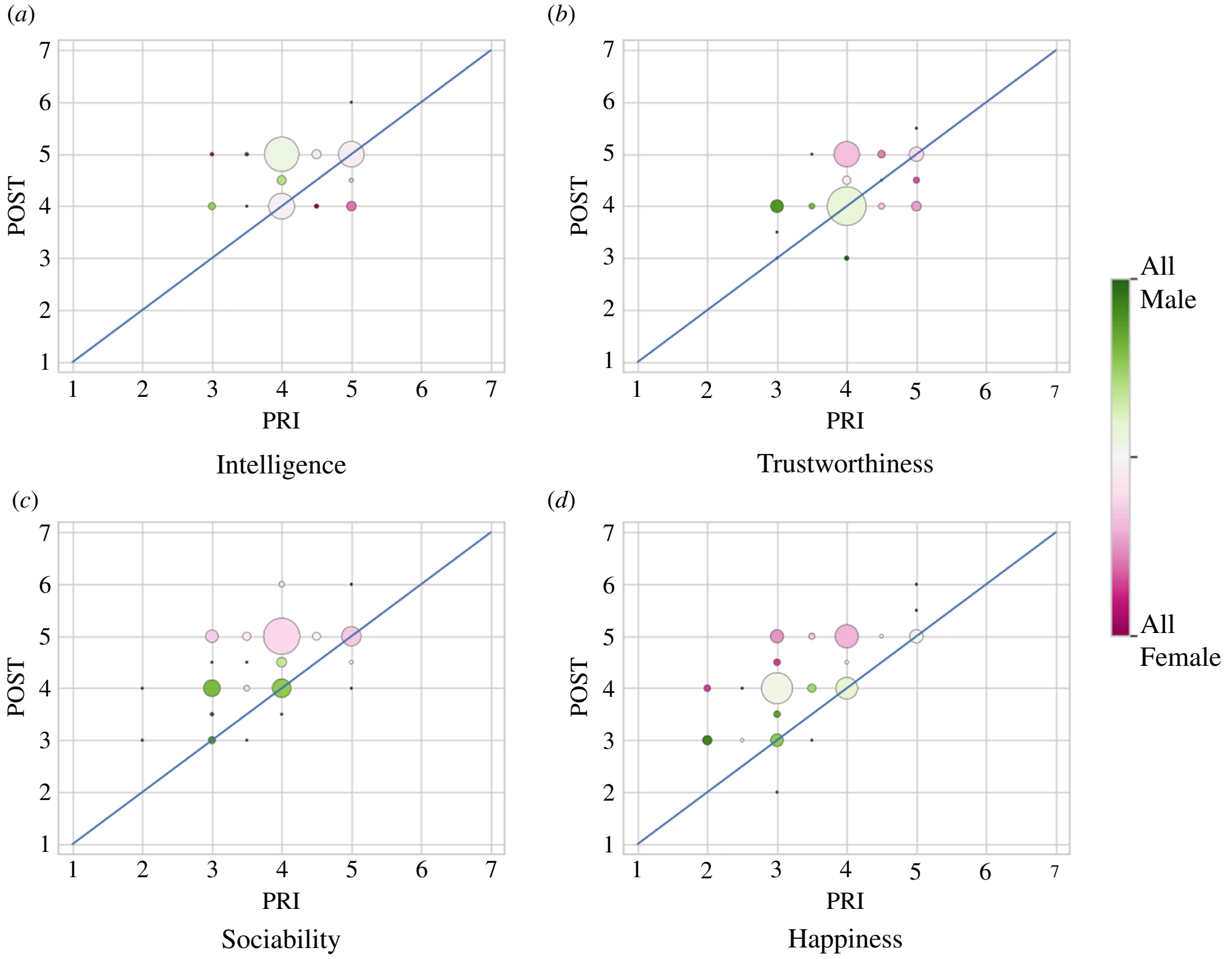

**Figure 13.** Visualization of the pairwise change in centralized scores of the dependent variables after applying the beauty filters. The *x*-axis represents the score an image received in the PRI dataset and the *y*-axis represents the score the corresponding image received in the POST dataset. The size of the circles is proportional to the number of images with the PRI and POST scores represented by the point and the colour of the circles represents the proportion of males and females at that point.

**Table 6.** One-sided Wilcoxon paired-rank tests ($W$) normalized over the number of samples ($n = 462$) comparing the median values for each of the dependent attributes in the original (PRI) and beautified (POST) faces. The same individuals were perceived as more intelligent, trustworthy, sociable and happy after beautification. *** denotes $p < 0.001$.

| | attractiveness | intelligence | trustworthiness | sociability | happiness |
|---|---|---|---|---|---|
| $W/n$ | 213.83*** | 63.15*** | 33.13*** | 123.49*** | 118.57*** |

# Appendix J. Model selection

Section 2.3 describes the impact of the participants' age and gender on the attractiveness halo effect by means of ordered stereotype models (OSM) [58] in conjunction with linear mixed models. In this section, we present the goodness of fit analyses that justified such model selection.

We evaluated 10 different models, depicted in figure 14, according to a taxonomy with three levels. The first level corresponds to the type of model (ordinal, such as the cumulative link model, or linear); the second level reflects whether the data is in the original 7-point Likert scale or in the new scales obtained by the OSMs. Furthermore, in the case of linear models, a third option is considered where the number of points in the scale is given by the OSMs yet the points are equidistant; the third level describes whether the raters were considered as fixed effects (FE; dashed line) or random effects (RE; solid line).

The 10 models were evaluated using the AIC [120] and BIC [121] on the data from the PRI and POST sets separately as reflected in tables 7–10. Note that the AIC and BIC are sensitive to sample size, which in our case is $N = 27\,480$ data points. Thus, the values presented in the tables are divided by a factor of $10^3$. The best fitting model corresponds to the lowest AIC/BIC values, which are marked in bold in the tables.

**Table 7.** AIC/$10^3$ on the PRI set for all variables and model variations.

| | $M1_{FE}$ | $M1_{RE}$ | $M2_{FE}$ | $M2_{RE}$ | $M3_{FE}$ | $M3_{RE}$ | $M4_{FE}$ | $M4_{RE}$ | $M5_{FE}$ | $M5_{RE}$ |
|---|---|---|---|---|---|---|---|---|---|---|
| attractiveness | 45.039 | 43.835 | 44.455 | **43.255** | 45.799 | 44.176 | 45.497 | 43.861 | 45.181 | 43.835 |
| intelligence | 36.035 | 35.184 | 29.349 | 28.482 | 36.213 | 35.260 | 31.771 | 31.064 | 22.618 | **21.937** |
| trustworthiness | 37.225 | 36.004 | 36.601 | 35.373 | 37.666 | 36.110 | 37.268 | 35.728 | 35.268 | **33.898** |
| sociability | 39.107 | 38.547 | 38.555 | 37.990 | 39.368 | 38.637 | 39.040 | 38.311 | 36.040 | **35.279** |
| happiness | 40.272 | 39.460 | 40.283 | 39.470 | 40.623 | 39.586 | 40.637 | 39.598 | 31.273 | **30.187** |

**Table 8.** AIC/$10^3$ on the POST set for all variables and model variations.

| | $M1_{FE}$ | $M1_{RE}$ | $M2_{FE}$ | $M2_{RE}$ | $M3_{FE}$ | $M3_{RE}$ | $M4_{FE}$ | $M4_{RE}$ | $M5_{FE}$ | $M5_{RE}$ |
|---|---|---|---|---|---|---|---|---|---|---|
| attractiveness | 42.537 | 41.757 | 38.774 | **38.057** | 44.223 | 43.157 | 40.652 | 39.689 | 41.731 | 40.696 |
| intelligence | 34.622 | 33.270 | 33.544 | 32.234 | 34.825 | 33.204 | 34.174 | 32.661 | 33.642 | **32.156** |
| trustworthiness | 36.751 | 35.421 | 30.366 | **29.134** | 36.966 | 35.317 | 32.435 | 31.238 | 31.366 | 30.004 |
| sociability | 37.806 | 37.030 | 37.522 | 36.758 | 38.131 | 37.300 | 37.835 | 37.025 | 34.715 | **34.092** |
| happiness | 39.198 | 38.610 | 39.275 | 38.683 | 39.433 | 38.710 | 39.533 | 38.802 | 36.100 | **35.328** |

**Table 9.** BIC/$10^3$ on the PRI set for all variables and model variations.

| | $M1_{FE}$ | $M1_{RE}$ | $M2_{FE}$ | $M2_{RE}$ | $M3_{FE}$ | $M3_{RE}$ | $M4_{FE}$ | $M4_{RE}$ | $M5_{FE}$ | $M5_{RE}$ |
|---|---|---|---|---|---|---|---|---|---|---|
| attractiveness | 45.129 | 43.933 | 44.538 | **43.345** | 45.859 | 44.244 | 45.558 | 43.929 | 45.241 | 43.902 |
| intelligence | 36.170 | 35.327 | 29.462 | 28.603 | 36.280 | 35.336 | 31.839 | 31.139 | 22.686 | **22.012** |
| trustworthiness | 37.361 | 36.147 | 36.721 | 35.501 | 37.734 | 36.186 | 37.336 | 35.803 | 35.336 | **33.973** |
| sociability | 39.242 | 38.690 | 38.675 | 38.118 | 39.436 | 38.712 | 39.108 | 38.386 | 36.108 | **35.354** |
| happiness | 40.408 | 39.603 | 40.411 | 39.606 | 40.691 | 39.661 | 40.704 | 39.673 | 31.341 | **30.262** |

**Table 10.** BIC/$10^3$ on the POST set for all variables and model variations.

| | $M1_{FE}$ | $M1_{RE}$ | $M2_{FE}$ | $M2_{RE}$ | $M3_{FE}$ | $M3_{RE}$ | $M4_{FE}$ | $M4_{RE}$ | $M5_{FE}$ | $M5_{RE}$ |
|---|---|---|---|---|---|---|---|---|---|---|
| attractiveness | 42.628 | 41.855 | 38.849 | **38.140** | 44.283 | 43.224 | 40.712 | 39.756 | 41.791 | 40.764 |
| intelligence | 34.758 | 33.413 | 33.657 | 32.354 | 34.893 | 33.279 | 34.242 | 32.736 | 33.709 | **32.232** |
| trustworthiness | 36.887 | 35.564 | 30.479 | **29.255** | 37.034 | 35.392 | 32.503 | 31.314 | 31.434 | 30.079 |
| sociability | 37.941 | 37.173 | 37.635 | 36.878 | 38.199 | 37.375 | 37.902 | 37.100 | 34.783 | **34.168** |
| happiness | 39.333 | 38.753 | 39.396 | 38.811 | 39.501 | 38.785 | 39.601 | 38.877 | 36.167 | **35.403** |

Based on these results, we opted for $M5_{RE}$, i.e. a linear mixed model on the OSM-based rescaled data with the raters as random effects. In addition to being the best performing model in most cases, linear mixed models are significantly more interpretable than ordinal models [119].

## Appendix K. Evaluation of the saturation effect in the halo effect

As discussed in §2.4, we observe a saturation in the relationship between attractiveness and some of the dependent variables, namely intelligence and trustworthiness (see figure 5). As an initial test of this hypothesis, table 11 summarizes the results of Wilcoxon paired-rank tests on the centralized scores of the images with perceived attractiveness scores greater than or equal to 5 before beautification (highly attractive stimuli) compared with the remaining stimuli and the complete dataset. Differences in pairwise perceived intelligence and trustworthiness before and after beautification are not statistically significant for the 'highly attractive stimuli' whereas they are significant ($p < 0.001$) in the rest of the cases. This finding supports the saturation hypothesis for intelligence and trustworthiness.

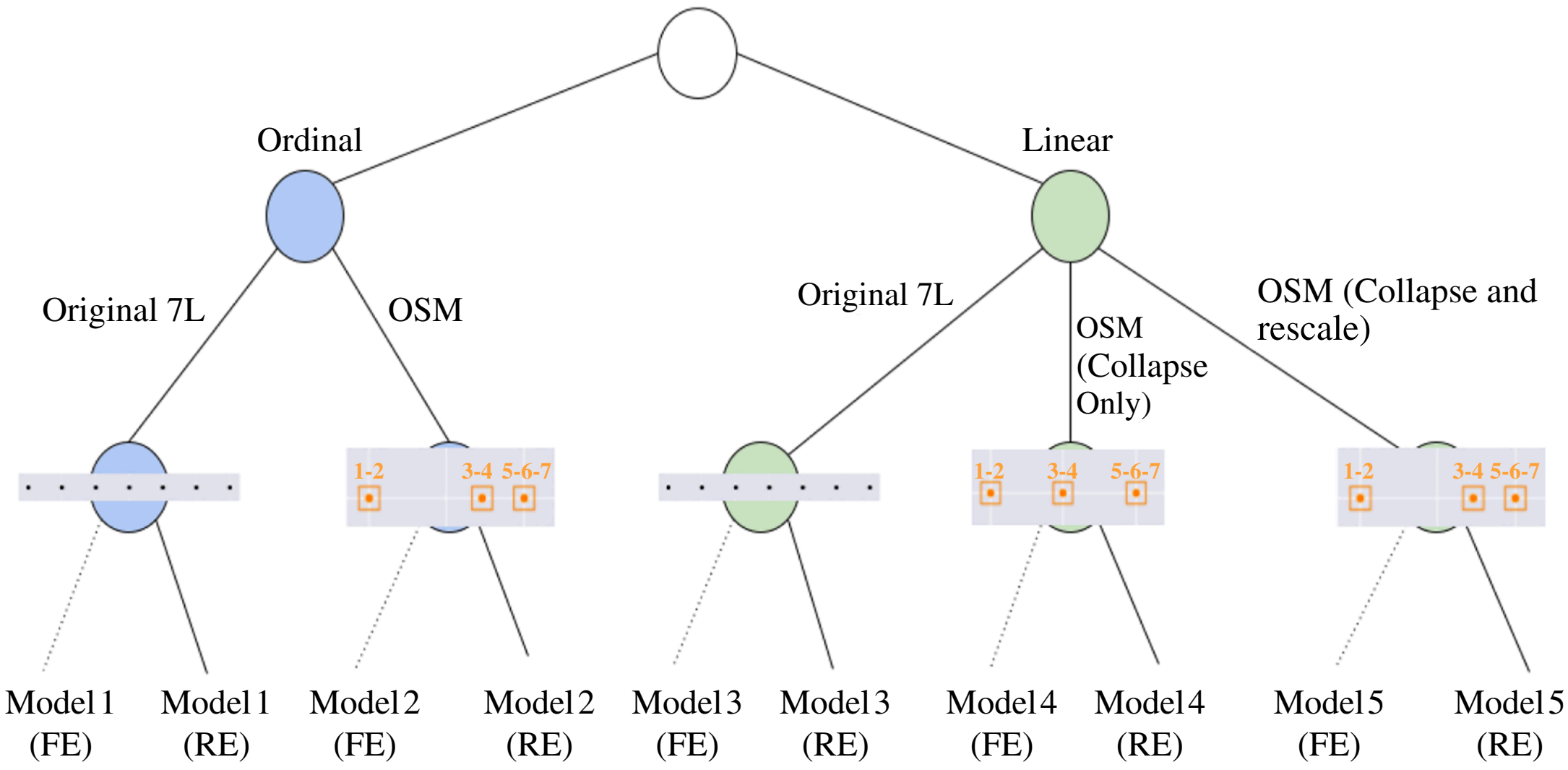

**Figure 14.** Three-level taxonomy of modelling choices that were evaluated for their goodness of fit, resulting in 10 different models.

**Table 11.** Normalized Wilcoxon paired-rank test statistic ($W/n$) for attractiveness and the four dependent attributes. An image is considered highly attractive if its centralized attractiveness score is ≥ 5 in the PRI dataset.

| dependent attribute (ω) | complete image set ($n$ = 462) | highly attractive stimuli ($n$ = 79) | remaining stimuli ($n$ = 383) |
|---|---|---|---|
| attractiveness | 213.83*** | 32.35*** | 182.12*** |
| **intelligence** | 63.15*** | **3.22** | 61.81*** |
| **trustworthiness** | 33.13*** | **6.71** | 26.31*** |
| sociability | 123.49*** | 10.86*** | 115.04*** |
| happiness | 118.57*** | 14.28*** | 105.28*** |

We further quantify the strength of the effect by means of two approaches:

## K.1. Method A: piece-wise linear fit

For each dependent variable, the data is divided in two halves according to the corresponding attractiveness ratings. A linear function ($\omega = m.Attrac + c$) is fitted to each set and the slopes ($m$) of the linear functions are compared to quantify the saturation effect as attractiveness increases, $Sat^A_\omega = m_{Upper} - m_{Lower}/m_{Lower} \times 100$, where $m_{Lower}$ and $m_{Upper}$ represent the slopes of the lines fit on the lower and upper part of the data, respectively. We report these findings in the first column of table 12. The strongest saturation effect is observed for intelligence followed by trustworthiness.

## K.2. Method B: fitting a log curve

A second approach to measure the strength of the saturation effect consists of fitting a log curve of the form $\omega = a \cdot log(attrac) + b$ and comparing the goodness of fit (by means of the AIC [120]) with a line of the form $\omega = a \cdot attrac + b$. Figure 15 depicts the log curves fit on the data in blue. To evaluate the strength of the saturation effect, we measure the percentage change in the AIC of both the log and the linear curves, $Sat^B_\omega = AIC_{Linear} - AIC_{Log}/AIC_{Log} \times 100$, where $AIC_{Linear}$ and $AIC_{Log}$ represent the AICs of the linear fit and log curve, respectively. Since a lower AIC indicates a better fit, the larger the value of $Sat^B_\omega$, the stronger the saturation effect. Again, the strongest saturation is observed for intelligence followed by trustworthiness.

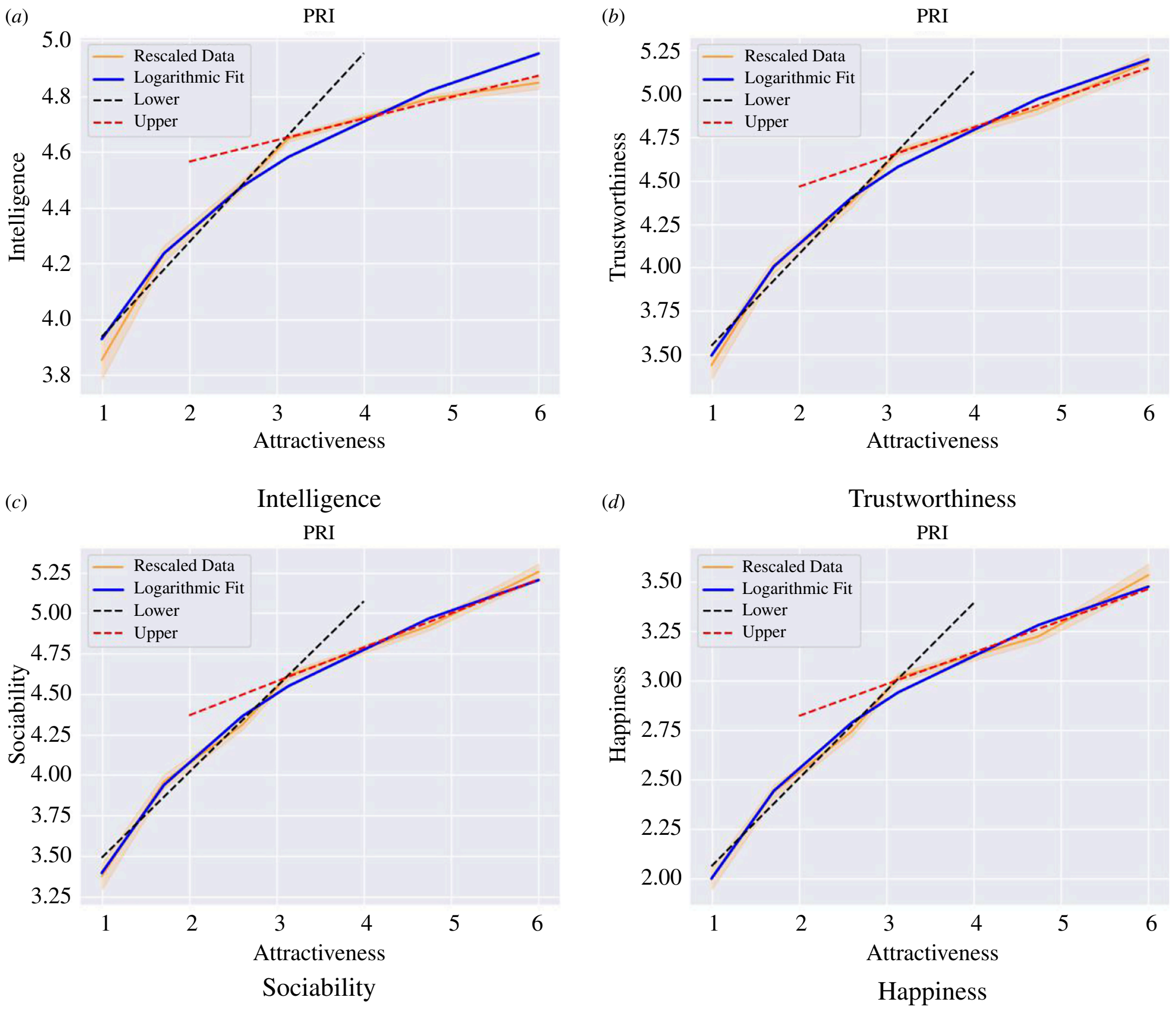

**Figure 15.** Relationship between attractiveness and the dependent attributes after rescaling the data on the PRI set. The yellow curve represents the rescaled data, the blue curve represents a logarithmic curve fit to the data and the dashed lines represent the best fit lines on the lower (black) and upper (red) half of the data. While all attributes show saturation to a degree, it is strongest for intelligence and trustworthiness. Note that the $y$-axis values for the different dependent attributes are not directly comparable since they were all rescaled independently using the OSMs.

**Table 12.** Evaluation of the strength of the saturation effect for different dependent variables using the methods described in appendix K. Note how the effect (difference between the values between the PRI and POST datasets) is strongest for intelligence followed by trustworthiness.

| dependent attribute ($\omega$) | method A (K) | | method B (K) | |
|---|---|---|---|---|
| | PRI | POST | PRI | POST |
| intelligence | −77.2 | −54.98 | 2.3 | 0.37 |
| trustworthiness | −67.50 | −68.88 | 1.08 | 0.68 |
| sociability | −60.22 | −45.26 | 0.83 | 0.35 |
| happiness | −63.83 | −49.35 | 0.91 | 0.43 |

# Appendix L. Linear mixed models including rater effects

The parameters for all the linear mixed models described in §4.5.2 can be found in the anonymized GitHub repository containing the code used to analyse the data. The model parameters can also be directly accessed at this link: https://tinyurl.com/modelParametersFile-ea. The parameters of the fixed-effects terms represent the slopes corresponding to each of the terms. Note that females (for gender of the image and the gender of the rater) are coded as 0. Thus, the beta values correspond to the slope for images of males and male raters. The impact of the gender of the image and rater and their interactions have been discussed in §2.3 and appendix H by computing the estimated marginal means instead of relying only on the $\beta$s presented here.

**Table 13.** Partial $R^2$ of each of the predictors in the linear mixed models described in appendix L. Note how attractiveness has the largest $R^2$ of all the variables, indicating that attractiveness best explains the variance of the dependent variables.

| dependent attribute ($\omega$) | | full model | attractiveness | image gender | image age | participant gender | participant age |
|---|---|---|---|---|---|---|---|
| attractiveness | PRI | 0.145 | X | $e^{-12}$ | 0.015 | 0 | 0.001 |
| | POST | 0.195 | X | $e^{-12}$ | 0.014 | $e^{-12}$ | 0 |
| intelligence | PRI | 0.150 | 0.138 | $e^{-12}$ | 0.001 | $e^{-9}$ | 0.001 |
| | POST | 0.098 | 0.083 | $e^{-12}$ | 0.007 | $e^{-11}$ | 0.007 |
| trustworthiness | PRI | 0.169 | 0.141 | 0 | 0.001 | 0 | $e^{-4}$ |
| | POST | 0.087 | 0.065 | 0 | 0.002 | 0 | 0.003 |
| sociability | PRI | 0.192 | 0.157 | $e^{-12}$ | $e^{-5}$ | $e^{-12}$ | 0.001 |
| | POST | 0.167 | 0.109 | 0 | $e^{-4}$ | 0 | 0 |
| happiness | PRI | 0.177 | 0.152 | 0 | $e^{-5}$ | $e^{-9}$ | 0.002 |
| | POST | 0.141 | 0.100 | $e^{-12}$ | $e^{-4}$ | $e^{-9}$ | 0.002 |

# Appendix M. Partial $R^2$ in the linear mixed models

In order to evaluate the importance of attractiveness in predicting the dependent variables when compared with other predictors, we compute the partial $R^2$s [134] of each of the predictors for all the linear mixed models. The results are summarized in table 13. Observe how attractiveness explains the largest part of the variance of all of the models.

In appendix C, we noted that the filters reduce perceptions of age of the stimuli. Given, however, that the partial $R^2$ associated with age is much lower than the partial $R^2$ for attractiveness, we conclude that it is the change in attractiveness driving the changed perceptions of the dependent variables and not the change in the perceived age of the subjects in the images.

# Appendix N. Computation of fractional change in estimated marginal means

This section describes in detail how the $y$-axis values of the plots in figure 4 were computed. The values are directly proportional to the EMM. However, the OSM provide different scales for each attribute in the PRI and POST datasets, which makes it hard to directly compare values computed on the PRI and POST scales. However, to understand the impact of the gender of the rater, it is sufficient to compare relative changes between image gender–rater pairs. Thus, setting the value of images of females rated by females as 0 enables a comparison of the relative changes between (rater, image) gender pairs. Thus, the fractional change values depicted on the $y$-axis of the graphs in figure 4 for each dependent attribute $\omega$ in the PRI and POST datasets were computed as

$$fractionalChange = \frac{EMM_{(i,j)} - EMM_{(f,f)}}{numLevels}, \quad \text{(N 1)}$$

where $EMM_{(f,f)}$ represents the estimated marginal mean value of images of females rated by female raters for the dependent attribute $\omega$ in the PRI or POST datasets. $EMM_{(i,j)}$ represents the corresponding EMM for every image gender–rater gender pair $(i, j)$ in the same setting and *numLevels* represents the number of levels on the rescaled version of the dependent attribute.

This computation of the fractional changes enables a visualization of (i) how different raters are impacted by the gender of the stimulus (i.e. differences between the blue and pink bars for each rater gender); (ii) how the gender of the stimuli impacts the perceptions provided by the raters (i.e. how different are the two pink (or blue) bars between male and female raters); and (iii) how different are these changes between the PRI and POST datasets for each dependent attribute, reflecting the impact of the beauty filters.